\documentclass[twocolumn,aps,showpacs]{revtex4}
\usepackage{graphicx}
\usepackage{amssymb}
\usepackage{subfigure}

\usepackage{bm}

\begin{document}

\title{Targeted mixing in an array of alternating vortices}

\author{R. Bachelard$^1$}
\author{T. Benzekri$^{1,2}$}
\author{C. Chandre$^1$}
\author{X. Leoncini$^{1,3}$}
\author{M. Vittot$^1$}

\affiliation{$^1$Centre de Physique Th\'{e}orique\footnote{Unit\'{e} Mixte de
Recherche (UMR 6207) du CNRS, et des universit\'{e}s Aix-Marseille
I, Aix-Marseille II et du Sud Toulon-Var. Laboratoire affili\'{e}
\`{a} la FRUMAM (FR 2291). Laboratoire de Recherche Conventionn\'e
du CEA (DSM-06-35).}, CNRS Luminy, Case 907, F-13288
Marseille cedex 9, France\\ $^2$ Universit\'e des Sciences et de Technologie HB, BP 32 EL Alia,
Bab Ezzouar Alger, Algeria\\ $^3$ PIIM, Universit\'e de Provence-CNRS, Centre Universitaire de Saint
J\'er\^ome, F-13397 Marseille, France}

\date{\today}

\begin{abstract}
Transport and mixing properties of passive particles advected by an
array of vortices are investigated. Starting from the integrable case,
it is shown that a special class of perturbations allows one to preserve
separatrices which act as effective transport barriers, while triggering
chaotic advection. In this setting, mixing within the two
dynamical barriers is enhanced while long range transport is prevented.
A numerical analysis of mixing properties depending on parameter values
is performed; regions for which optimal mixing is achieved are proposed.
Robustness of the targeted mixing properties regarding errors in the
applied perturbation are considered, as well as slip/no-slip boundary
conditions for the flow.
\end{abstract}


\maketitle

\section{Introduction}

Since its uncovering, chaotic advection \cite{Aref84,Aref90} has
drawn much attention as its consequences on both transport and mixing
properties of a given flow are fundamental. This phenomenon is intimately
related to Lagrangian chaos, and translates the fact that despite
the laminar character of a flow Lagrangian trajectories of a fluid or
passive particles may end up being chaotic. In this setting, transport
and mixing properties are drastically changed \cite{Ottino89,Ottino90,Zaslav91,Crisanti91}.
In chaotic regions of the flow, mixing induced by molecular diffusion
becomes often negligible in regards to the mixing induced by the dynamics.
Regarding transport properties, the triggering of chaotic advection
also plays a key role. Indeed, in contrast to the fully predictive
integrable situation, tackling transport properties of individual
passive particles is subject to sensitivity to initial conditions  and
implies a necessary probabilistic approach. This leads in some cases
to a diffusion equation with an enhanced diffusion coefficient when
compared to the molecular diffusion one \cite{Rechester80}, but also
to non-Gaussian properties of transport (see for instance\cite{Solomon94,Leoncini05}),
as for instance super-diffusive transport. One then often resorts
to modeling transport using a fractional diffusion equation \cite{Zaslavsky2002}.
All these properties have drawn much attention not only for its impact
on geophysical flows and magnetized plasmas \cite{Brown91,Behringer91,Chernikov90,Dupont98,Crisanti92,Carreras03,Annibaldi00,Castillo2004,Leoncini05},
but also because of the potential applications of its enhanced mixing
properties in chemical engineering and micro-fluidic devices \cite{Balasuriya05,Stroock02}.

Lagrangian chaos in two dimensional incompressible flows is triggered
generically when the flow becomes time-dependent\cite{Solomon88,Solomon03,Aref90,Willaime93,Camassa91}. The trajectories of fluid particles
or passive tracers are not confined on field lines and chaos appears.
For these type of flows the dynamics of fluid particles is Hamiltonian,
with the stream function acting as the Hamiltonian. The canonically
conjugate variables are the space variables, and the phase space is the two dimensional physical space. This
particularity allows direct visualization of phase space in experiments,
making it a test bed used to confront theoretical
results on Hamiltonian dynamics with experiments. More specifically,
passive tracer dynamics belong to the class of Hamiltonian systems with one and a half degrees
of freedom. General mixing and transport properties of
these systems are now well understood, especially when the time dependence
is periodic and Poincar\'e maps are computed to analyze phase space
topology. Typically the dynamics is not ergodic~: A chaotic sea
 surrounds various islands of quasi-periodic
dynamics. The anomalous transport properties and their multi-fractal
nature are then linked to the existence of islands and the phenomenon
of stickiness observed around them \cite{Kuznetsov2000,LKZ01}, while
mixing is enhanced in the chaotic sea but has to rely on molecular
diffusion in regular regions.

In most studies regarding this type of phenomena, the time dependent
perturbation is given a priori or self-generated. Transport and mixing
properties are thoroughly investigated and general laws are extrapolated
or the origin of phenomena explained (see \cite{Villermaux03,LKZ01,Vainchtein,Castiglione99}).
The influence of phase space topology and its understanding
is clear, and can be used to explain synchronization
phenomena \cite{Paoletti06}. However one is still somewhat constrained
by an a priori imposed time dependence. Recently, approaches of tailoring
specific perturbations in order to modify phase space have been proposed
\cite{Chandre05,Bachelard06} and a specific one
has been applied to an array of alternating vortices \cite{Benzekri06}.
Due to the strong influence of invariant tori forming regular islands
on global transport and mixing properties, acting on phase space topology
even locally (for instance, by building a transport barrier \cite{Chandre05,Benzekri06}
or by destroying regular islands \cite{Benzekri06}) can have strong
consequences.

In this paper we address the problem of targeted mixing, a work already
started in Ref.~\cite{Benzekri06}. We refer to targeted mixing as
the process of leveraging only one of the consequences of chaotic
advection, namely enhancing mixing, while containing particle transport
within a finite sub-domain of phase space. For this purpose we consider
the dynamics of passive particles in an array of alternating vortices
and tailor phase space in order to achieve the desired property. From
the experimental point of view, acting on phase space is
often not easy, because one has to act also on particle momenta in
general. However this may be less of a problem in the context of two-dimensional
incompressible flows due to the duality between physical space and
phase space. Moreover mixing within flows has a tremendous number
of applications. The primary interest in the flow of an array of alternating
vortices resides in the fact that being generated by quite a few hydrodynamic
instabilities, it may be considered as one of the founding bricks
of turbulence. This flow is easily accessible to experimentalists
for instance using magnetohydrodynamics techniques similar to Rayleigh-B\'{e}nard convection with a control
over the flow~\cite{Solomon88,Solomon03,Willaime93}.
As such, understanding its influence on the advection of passive or
active quantities is considered a necessary first step in order to
uncover the different mechanisms governing transport in general or
reaction-diffusion processes such as front propagation in turbulent
flows~\cite{Cencini03,Pao05a,Pao05b,Pocha06,Vergni05}. The stream function which
models an experiment in a channel
with slip boundary conditions writes: \begin{equation}
\Psi_{0}(x,y)=\sin x\sin y,\label{psi0}\end{equation}
 where the $x$-direction is the horizontal one along the channel
and the $y$-direction is the bounded vertical one. The dynamics given
by the stream function~(\ref{psi0}) is integrable and passive particles follow the
stream lines, no mixing occurs. In the experiments, a typical perturbation
$f(x,y,t)$ is introduced as a time dependent forcing in order to
trigger chaotic advection and then to study the resulting transport
and mixing properties. More precisely we consider perturbations which
modify the stream function as: \begin{equation}
\Psi_{c}(x,y,t)=\Psi_{0}(x+f,y).\label{psic}\end{equation}

We show how to identify the perturbations $f$ which preserve transport barriers and at the same time, enhance mixing properties.

 The paper is organized as follows: In Sec.~\ref{sec:Definition-and-short}
we recall some basic notions of passive scalar dynamics in two dimensional
incompressible flows. Then we give a short review of possible Hamiltonian
control techniques we use in order to tailor a perturbation best suited
for our needs. In Sec.~\ref{sec:Achieveing-targeted-mixing}, we
apply these techniques to flows modeled by the stream function (\ref{psi0}).
First we derive the perturbation needed in order to build virtual
barriers along the channel and thus limit transport within only a
small region of phase space. Then among all possible perturbations,
we define criteria needed to achieve good mixing within the two barriers
and analyze for which perturbations and parameters these criteria
are satisfied. Then, we consider the robustness of the proposed perturbation
with respect to a simpler time dependence, slip boundary conditions,
three dimensional effects and molecular diffusivity. After showing
the efficiency of such perturbations in enhancing mixing, we analyze
the set of parameters for which efficient mixing is expected using
the residue method.

\section{A strategy for mixing inside a cell\label{sec:Definition-and-short}}

The term advection by definition, relates to the action of being moved
by and with a flow. In mathematical terms, this translates in the
general equation for a passive tracer:\begin{equation}
\dot{{\bf {r}}}={\bf v}({\bf r},t)\:,\label{advecbasic}\end{equation}
 where ${\bf r}$ locates the passive tracer in space, ${\bf v}$
corresponds to the velocity field, and $\dot{{\bf r}}$ corresponds
to the time derivative of ${\bf r}$. In the case of two dimensional
incompressible flows, the study of the field can be described by a
scalar function, i.e the stream function $\psi({\bf r},t)$. The velocity
field is then obtained by ${\bf v}=\mathrm{curl}(\psi\;\hat{{\bf z}})$,
where $\hat{{\bf z}}$ is the unit vector normal to the flow. Equation~(\ref{advecbasic})
is rewritten using the stream function and exhibits a Hamiltonian
structure for the flow~:\begin{equation}
\dot{x}=-\frac{\partial\Psi}{\partial y},\hspace{1.2cm}\dot{y}=\frac{\partial\Psi}{\partial x}\:,\label{eq:Hamilton_advec}\end{equation}
 where $(x,y)$ corresponds to the coordinates of the tracer on the
plane. The space variables $(x,y)$ are canonically conjugate for
the stream function $\Psi$ which acts as the Hamiltonian of the system.
Hence the phase space is formally the two dimensional physical space
(with the addition of time).

\begin{figure}
\begin{centering}\includegraphics[width=7cm]{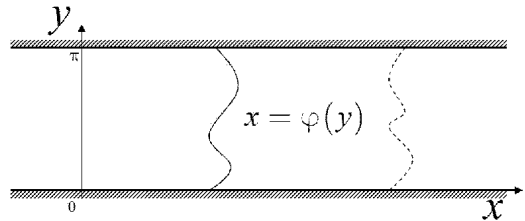} \par\end{centering}
\caption{\label{fig:canal} Schematic representation of the channel with a
transport barrier located at $x=\varphi(y)$.}
\end{figure}

In this section, our aim is to briefly recall some Hamiltonian techniques
allowing to some extent, to tailor phase space by adjusting appropriately
the perturbation $f$ as in Eq.~(\ref{psic}) and its parameters.
These techniques will be subsequently used to achieve targeted mixing
for the considered flows, namely construct a perturbation in order
to obtain optimal chaotic mixing in a localized region of phase space.
First we construct a family of perturbations which create transport
barriers, then we use the residue method~\cite{Bachelard06} in order
to find optimal mixing regimes in parameter space after the barriers
have been created.

\subsection{Constructing a perturbation with a barrier\label{sec:Constructing barrier}}

We consider a generic time-independent stream function (Hamiltonian)
$\Psi_{0}(x,y)$ describing
a fluid in a two-dimensional channel of height $\pi$, i.e.\ $(x,y)\in\mathbb{R}\times[0,\pi]$
(see Fig.~\ref{fig:canal}). We assume that there exists an invariant
curve which prevents the advection of tracers from the left to the
right of that curve. We denote the equation of this curve $x=\varphi(y)$.
The invariance condition {[}$\dot{x}=\varphi'(y)\dot{y}$ if $x=\varphi(y)$]
of tracers on this curve translates into~: \[
\Psi_{0}\left(\varphi(y),y\right)=constant,\]
 for all $y\in[0,\pi]$ by using Eqs.~(\ref{eq:Hamilton_advec}).
In the case of Eq.~(\ref{psi0}), this transport barrier is a heteroclinic
connection between two hyperbolic fixed points located at $y=0$ and
$y=\pi$.

In order to enhance mixing, the system is perturbed by a time-dependent
forcing (depending also on $x$ and $y$ in general). As a
consequence, its dynamics is generically no longer integrable, and
chaotic trajectories fills portions of the channel (these are parts
where mixing occurs). However, as a side effect, the transport barrier
is broken and trajectories start to diffuse along the channel (in
the $x$-direction).

In the following, we propose to design a time-dependent forcing of
the flow described by $\Psi_{0}$ which preserves the transport barrier
as well as the chaotic mixing. The aim is to find an appropriate forcing
which preserves a bounded domain of the channel with an enhanced chaotic
mixing inside. This domain can be bounded by two dynamical barriers as mentioned
above.

A main practical requirement is that the forcing should be as simple
as possible to be implemented. Consequently, we start by investigating
perturbations which only depends on $y$ and $t$. Furthermore, for
practical reasons, we restrict the perturbations to time-periodic
ones, and the period is chosen as $2\pi$ without loss of generality.
More precisely, we search for a perturbation $f(y,t)$ which modifies
the stream function $\Psi_{0}$ into~: \begin{equation}
\Psi_{c}(x,y,t)=\Psi_{0}\left[x+f(y,t),y\right].\label{eqn:phic}\end{equation}
 In order to simplify the computations, we perform a generic translation
in $x$ (which corresponds to a canonical transformation in the Hamiltonian
setting)~: \begin{eqnarray*}
\tilde{x} & = & x+\partial_{y}\beta(y,t),\\
\tilde{y} & = & y,\end{eqnarray*}
 where $\beta$ is a function to be specified later. In the new variables
$\tilde{x}$ and $\tilde{y}$, the dynamics is described by the stream
function~: \[
\tilde{\Psi}_{c}(\tilde{x},\tilde{y},t)=\Psi_{0}\left[\tilde{x}+f(\tilde{y},t)-\partial_{y}\beta(\tilde{y},t),\tilde{y}\right]-\partial_{t}\beta(\tilde{y},t).\]
 In order to have an invariant curve acting as a barrier in the channel,
the perturbation $f$ is such that the stream function evaluated at
$\tilde{x}=\varphi(\tilde{y})$ is only a function of time, since
the barrier is time-independent (in the moving frame). For simplicity,
we consider solutions for which the stream function vanishes~: \begin{equation}
\Psi_{0}\left[\varphi(y)+f(y,t)-\partial_{y}\beta(y,t),y\right]-\partial_{t}\beta(y,t)=0,\label{eqn:cond}\end{equation}
 which is a single equation with two unknown functions, $\beta$ and
$f$. In principle, there are an infinite set of solutions. Depending
on other requirements, some solutions are more appropriate than others.
In the following, we choose \[
f(y,t)+\varphi(y)-\partial_{y}\beta(y,t)=\Phi(t),\]
 where $\Phi(t)$ is any function of $t$. Equation~(\ref{eqn:cond})
implies \[
\partial_{t}\beta(y,t)=\Psi_{0}\left[\Phi(t),y\right].\]
 This equation has solutions provided that the mean value of $\Psi_{0}\left[\Phi(t),y\right]$
with respect to time vanishes for all $y$. This guides the choice
for the function $\Phi$. A possible solution for $\beta$ is \[
\beta(y,t)=\Gamma\Psi_{0}\left[\Phi(t),y\right],\]
 where \[
\Gamma v(y,t)=\sum_{k\neq0}\frac{v_{k}(y)}{ik}\mathrm{e}^{ikt},\]

for $v=\sum_{k}v_{k}(y)\mathrm{e}^{ikt}$. The perturbation $f$ is
given by \begin{equation}
f(y,t)=\Phi(t)-\varphi(y)+\Gamma\partial_{y}\Psi_{0}\left[\Phi(t),y\right].\label{pert}\end{equation}
 The stream function $\Psi_{c}$ given by Eq.~(\ref{eqn:phic}) has
an invariant curve whose equation is \begin{equation}
x=\varphi(y)-\Gamma\partial_{y}\Psi_{0}\left(\Phi(t),y\right).\label{barrier}\end{equation}

There exist many more solutions $f$ than the one indicated here. These solutions are obtained by using more complex translation functions
$\beta$. However, they contain a lot more Fourier modes in $y$, which make
them more difficult to implement for the cases we consider in Sec.~\ref{sec:Achieveing-targeted-mixing}.
We notice that the perturbation $f$ given by Eq.~(\ref{pert}) as
well as the equation of the transport barrier~(\ref{barrier}) are
parameterized by a time-periodic function $\Phi(t)$. This function
$\Phi$ is in general parameterized by essentially two parameters,
its frequency and its amplitude. In parameter space, the dynamics
characterized by the stream function $\Psi_{c}$ exhibit drastically
different behaviors depending on the values of parameters. In what
follows, we use periodic orbits to determine the regions in parameter
space where complete mixing inside the transport barriers occurs.
This method is briefly explained in the next section.

\subsection{Periodic orbit analysis\label{sec:residus}}

The dynamics of the perturbed flow as the one given by the stream
function $\Psi_{c}$ can be investigated by looking at the (linear)
stability of specific periodic orbits, using indicators such as the
\textit{residue}. In order to {}``control'' the flow, one can monitor
the residues by varying the parameters of the system (like the frequency
and the amplitude of the forcing), until specific bifurcations occur.
In particular, the residue method allows one to predict the break-up
(or creation) of invariant tori, which entails an enhancement (or
reduction) of chaotic mixing.

We consider a Hamiltonian flow with 1.5 degrees of freedom which depends
on a set of parameters ${\bm\lambda}\in\mathbb{R}^{m}$~:

\[
\dot{{\bf r}}={\mathbb{J}}\nabla_{{\bf r}}\Psi({\bf r},t;{\bm\lambda}),\]
 where ${\bf r}=(x,y)$ and ${\mathbb{J}}=\left(\begin{array}{cc}
0 & -1\\
1 & 0\end{array}\right)$. In order to analyze the linear stability properties of the associated
periodic orbits, we also consider the Jacobian $J^{t}({\bf r})$ which
evolves according to the tangent flow written as~\cite{chaosbook}~:
\begin{eqnarray}
\frac{d}{dt}{J^{t}}({\bf r})={\mathbb{J}}\nabla_{{\bf r}}^{2}\Psi({\bf r},t;{\bm\lambda})J^{t},\label{tgt_flow}\end{eqnarray}
 where $J^{0}$ is the two-dimensional identity matrix and $\nabla_{{\bf r}}^{2}\Psi$
is the Hessian matrix (composed of second derivatives of $\Psi$ with
respect to its canonical variables). For a given periodic orbit with
period $T$, the linear stability properties are given by the spectrum
of the two-dimensional monodromy matrix $J^{T}$. These properties
can be synthetically captured in Greene's definition \cite{Greene79,Mackay92} of the residue~:
\[
R=\frac{2-\mbox{tr}J^{T}}{4},\]
 since $\mbox{det }J^{T}=1$. In particular, if $R\in]0,1[$, the
periodic orbit is elliptic; if $R<0$ or $R>1$ it is hyperbolic;
and if $R=0$ and $R=1$, it is parabolic.

The residues of a set of well selected periodic orbits provide - through
linear stability analysis - information to detect the enhancement
as well as the reduction of chaotic mixing~\cite{Cary86,Bachelard06}.
The actual change of dynamics has to be checked a posteriori by a
nonlinear stability analysis (a Poincar\'{e} section for instance).
The residues are used to discard regions in parameter space where
large elliptic islands are present.

For this purpose, an alternative strategy is to use a {}``brute force''
method and scan a whole range of physically relevant parameters, analyze
transport and mixing properties (by a Poincar\'{e} map inspection
for instance) and conclude on a domain of parameters where optimal
mixing is achieved. Though it would be a complete
analysis, this strategy is not reasonable to adopt because of the
high number of cases to consider and also the computer time it takes
to analyze a single case.

In order to circumvent these difficulties we choose to consider the
residue method described in this section and follow in parameter space
the stability of a well selected set of periodic orbits. Indeed due
to the Hamiltonian nature of passive particles, one can expect a direct
correspondence between non-mixing regions in physical space and islands
of stability in phase space. The higher the period of the island the
smaller is its size, hence by following the linear stability of elliptic
periodic orbits with small period in parameter space, one should be
able to define a potential optimal mixing region for which these orbits
are unstable and associated with a mixing enhancement. Once this set
of main periodic orbits has been defined by close inspection of several
situations, the mixing region is obtained by looking at the bifurcation
curves in parameter space, e.g., the set of parameters such that the
residue is equal to one~\cite{Bachelard06}. We will show in the
next section how to combine the creation of transport barriers and
the residue method in a particular example of stream function given
by Eq.~(\ref{psi0}).

\section{Achieving targeted mixing in an array of vortices\label{sec:Achieveing-targeted-mixing} }

\begin{figure}[h!]
\begin{centering}

\includegraphics[width=4cm,height=4cm]{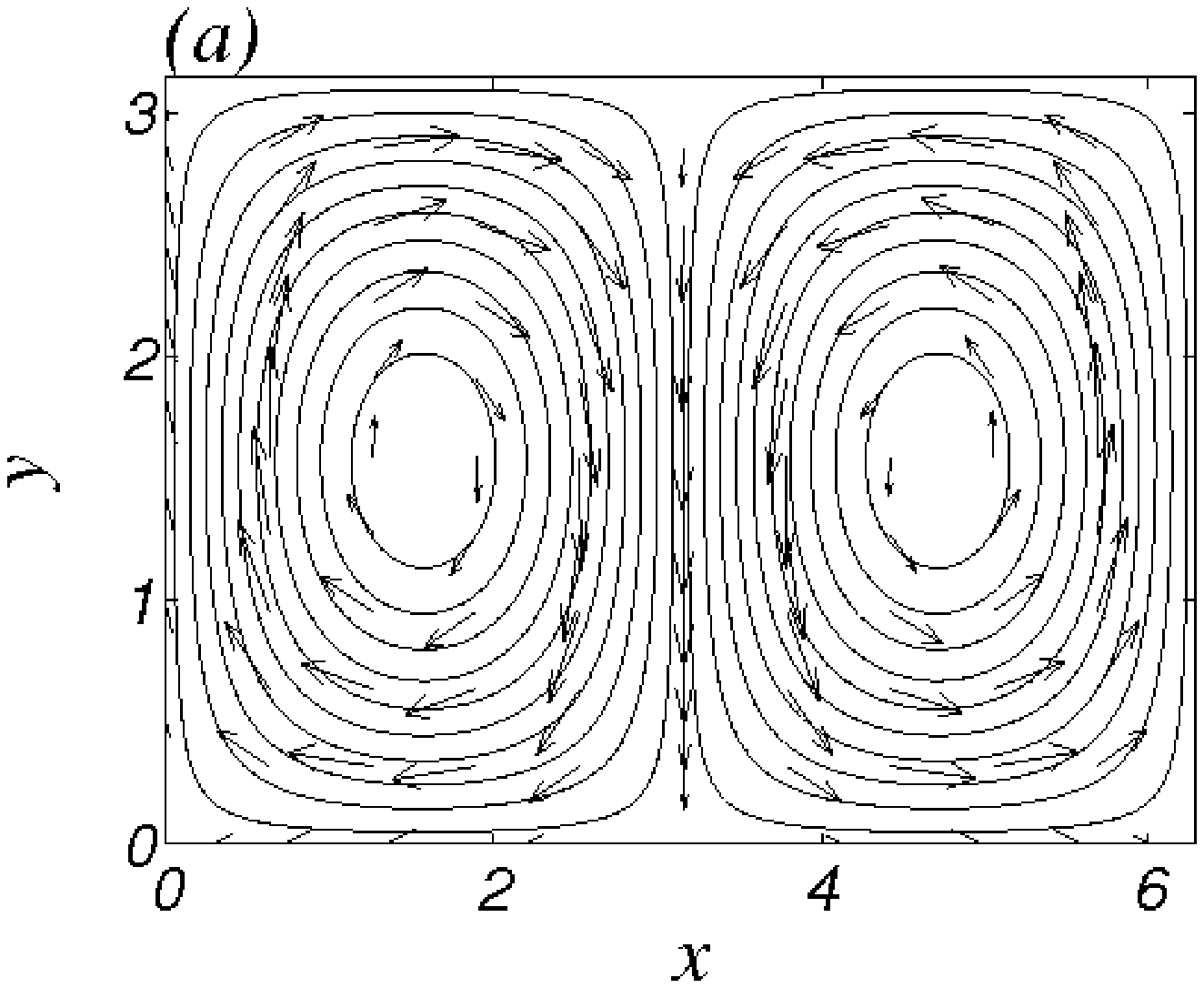}  \includegraphics[width=4cm,height=4cm]{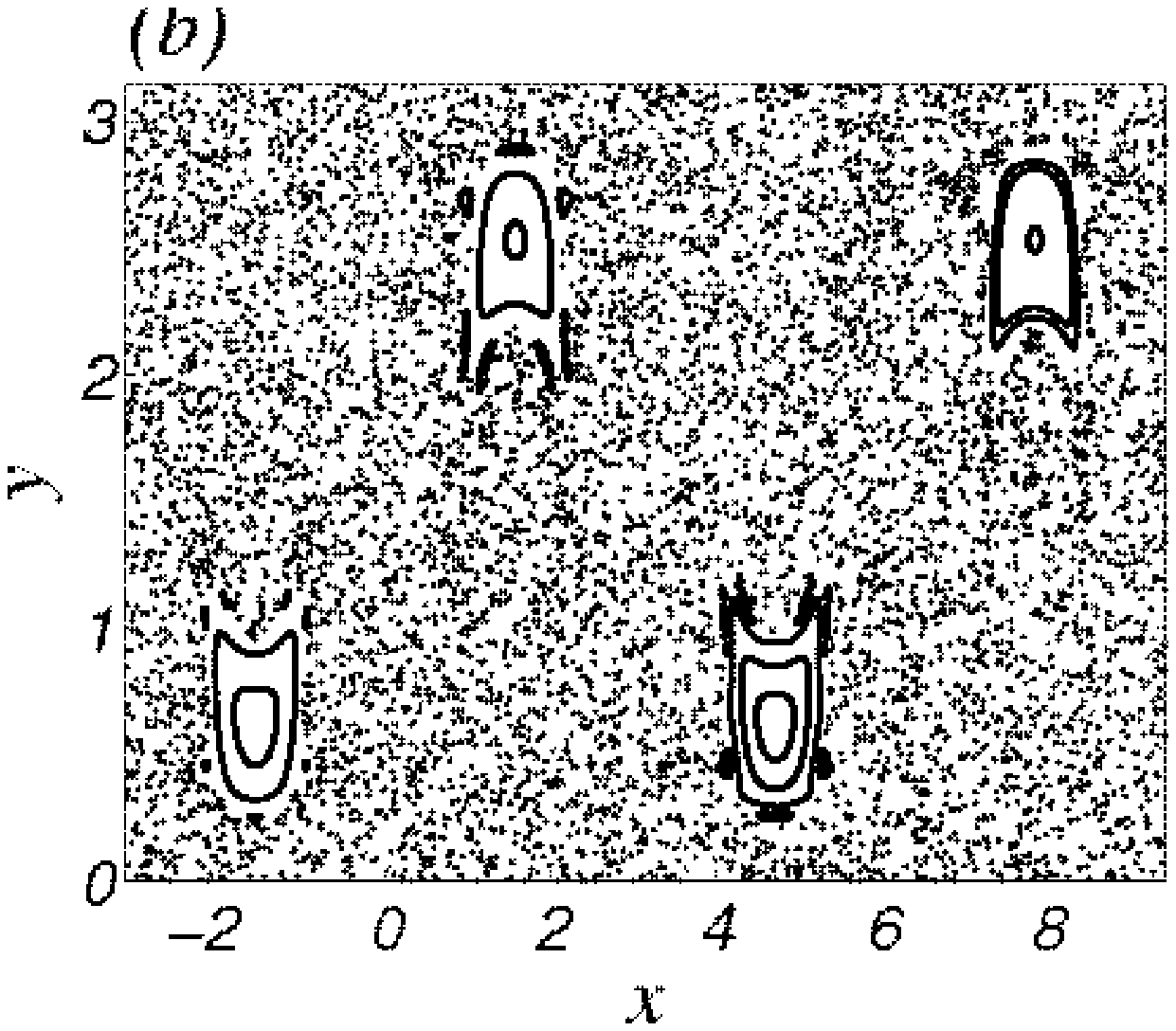}

\par\end{centering}
\caption{$(a)$ Streamlines at $t=0$ and and $(b)$ Poincar\'{e} section
of the stream function~(\ref{courant}). The parameters are $\omega=1$
and $\epsilon=0.8$.\label{Fignmix} }
\end{figure}

The stream function given by Eq.~(\ref{psi0}) models a cellular
flow consisting of alternating vortices. If we restrict ourselves
to $y\in[0,\pi]$, we have a channel of alternating vortices with
slip boundary conditions. From the Hamiltonian perspective, the advection
of passive tracers is given by Eq.~(\ref{eq:Hamilton_advec}). Since
the flow is steady, trajectories coincide with the fluid streamlines
depicted in Fig.~\ref{Fignmix}$(a)$.

Few basic facts explain the structures of the dynamics~: Boundary
conditions given in $y=0$ and $y=\pi$ constitute invariant curves.
The system is $2\pi$-periodic in the $x$-direction, i.e.\
along the channel. The system has hyperbolic fixed points located
at $x=m\pi$ for $m\in\mathbb{Z}$ and $y=0$ or $y=\pi$. These points
are joined by vertical heteroclinic connections for which the stable
and unstable manifolds coincide corresponding to roll boundaries,
at the origin of the cellular structure of the flow.

In order to obtain chaotic advection for two dimensional flows, one
has to perturb the flow by a time-dependent forcing. For example,
one can periodically force the roll patterns to oscillate in the $x$-direction
\cite{Solomon88}, in which case the stream function reads~:

\begin{equation}
\Psi_{1}(x,y,t)=\sin(x+\epsilon\sin\omega t)\sin y,\label{courant}\end{equation}
 where $f(x,y,t)=\epsilon\sin\omega t$ acts as a perturbation. The
parameter $\epsilon$ and $\omega$ are respectively the amplitude
and the angular frequency of these lateral oscillations.
The barriers (heteroclinic connections between the
hyperbolic periodic orbits) are broken under the perturbation $f$,
which cause the passive particles to undergo chaotic advection along
the channel~\cite{Wiggins92}.

The streamlines corresponding to the stream function $\Psi_{1}$ given
by Eq.~(\ref{courant}) are depicted in Fig.~\ref{Fignmix}$(a)$.
We observe that the structures are essentially the same as those
of the stream function $\Psi_{0}$ given by Eq.~(\ref{psi0}). However,
the periodic forcing now drives back and forth the roll patterns in
the $x$ direction with a period $2\pi/\omega$. The surfaces $y=0$
and $y=\pi$ are left invariant by the perturbation and the hyperbolic
orbits persist on these surfaces.
The Poincar\'{e} section for $\omega=1$ and $\epsilon=0.8$ (see
Fig.~\ref{Fignmix} $(b)$) shows how passive particles are spreading
along the channel. Particle transport from roll to roll is
greatly enhanced. However, an unmixed area characterized by regular
trajectories is still present at the center of each vortex : It is
composed of invariant tori of the dynamics. Consequently, though diffusion
has appeared in the system due to the time-dependent perturbation,
some regular patterns are persistent. Note that increasing the amplitude
$\epsilon$ does not make them disappear; in particular, in the limit
of large $\epsilon$ (with $\omega$ fixed), the system will be integrable as well.

\subsection{Building transport barriers}

In what follows, we propose to adjust the time periodic forcing as explained in Sec.~\ref{sec:Constructing barrier}
First we notice that $\Psi_{0}(m\pi,y)=0$ for any $y\in[0,\pi]$ and
all $m\in\mathbb{Z}$. We apply the construction of the perturbation with $\varphi(y)=0$ (or equivalently $\varphi(y)=2m\pi$). In Eq.~(\ref{pert}),
we choose $\Phi(t)=\epsilon\sin\omega t$ which gives the perturbation~:
\begin{equation}
f(y,t)=\epsilon\sin\omega t+\omega^{-1}\cos yC_{\epsilon}(\omega t)\:,\label{control}\end{equation}
 where \begin{eqnarray}
C_{\epsilon}(t) & = & \Gamma\sin(\epsilon\sin t)\nonumber \\
 & = & -2\sum_{n\geq0}\frac{1}{2n+1}\mathcal{J}_{2n+1}(\epsilon)\cos(2n+1)t,\label{serbessl}\end{eqnarray}
 and $\mathcal{J}_{n}$ (for $n\in\mathbb{N}$) are Bessel functions
of the first kind. In the numerics, we truncate the sum~(\ref{serbessl}) to five modes. Very similar results are obtained with a higher number of modes.

We notice that since the stream function is still $2\pi$ periodic
in the $x$-direction by construction, the invariant surface which
has been created around $x=0$ is also translated around $2m\pi$,
$m\in\mathbb{Z}$. The equations of these transport barriers along
the $x$-direction are~: \begin{equation}
x=2m\pi-\omega^{-1}\cos yC_{\epsilon}(\omega t),\label{eqbar}\end{equation}
 for all $m\in\mathbb{Z}$. Notice that these barriers exist for arbitrary
values of $\omega$ and $\epsilon$. Each of these barriers are heteroclinic
connections between two hyperbolic periodic orbits: \begin{eqnarray*}
x(t) & = & 2m\pi-\omega^{-1}C_{\epsilon}(\omega t)\quad\mbox{at}\quad y=0,\\
x(t) & = & 2m\pi+\omega^{-1}C_{\epsilon}(\omega t)\quad\mbox{at}\quad y=\pi,\end{eqnarray*}
 which move in opposite directions.
Furthermore, the top and bottom boundaries of the channel remain invariant.
This comes from the fact that the perturbation given by Eq.~(\ref{control})
is only applied in the $x$ term of $\Psi_0$.

\begin{figure}[h!]
\begin{centering}\includegraphics[width=4cm,height=4cm]{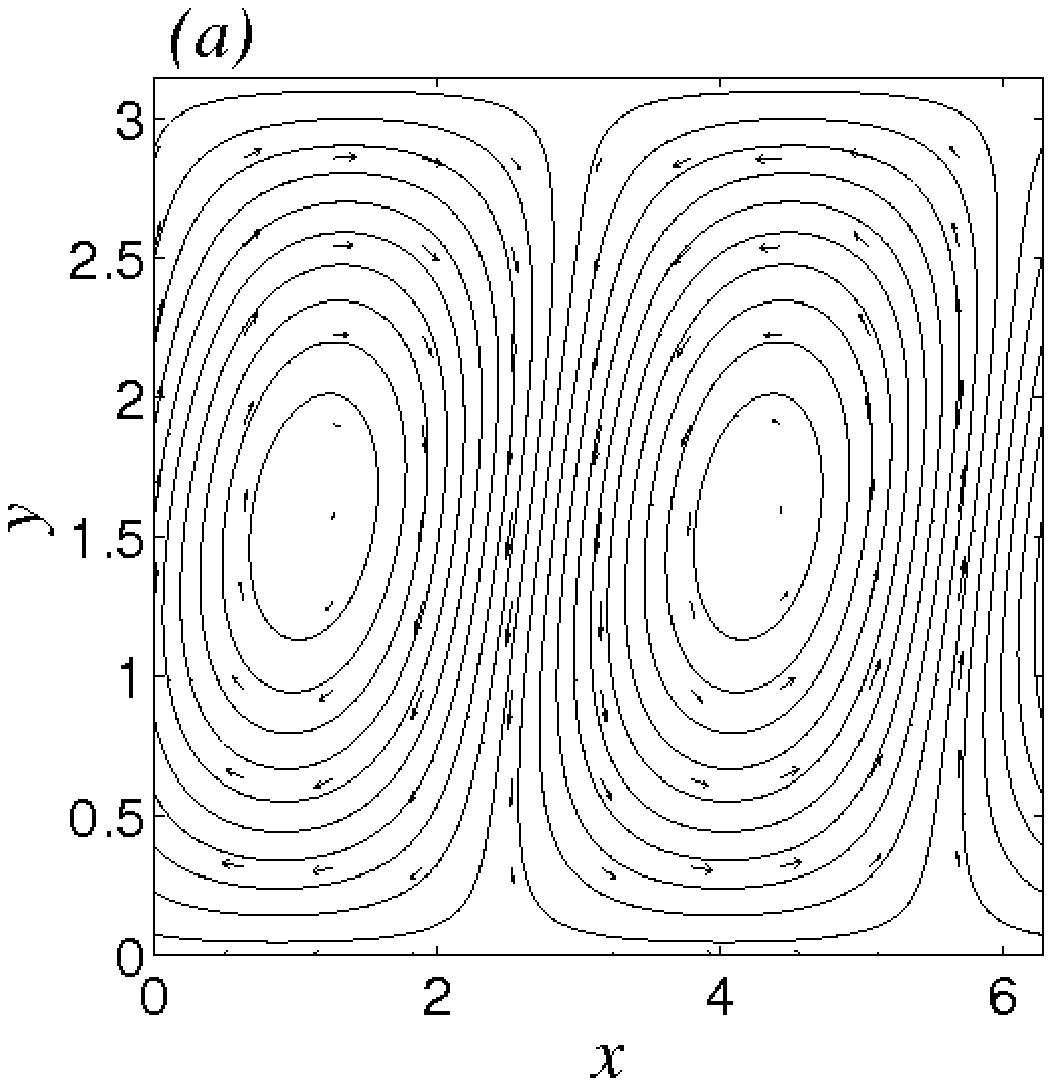} \includegraphics[width=4cm,height=4cm]{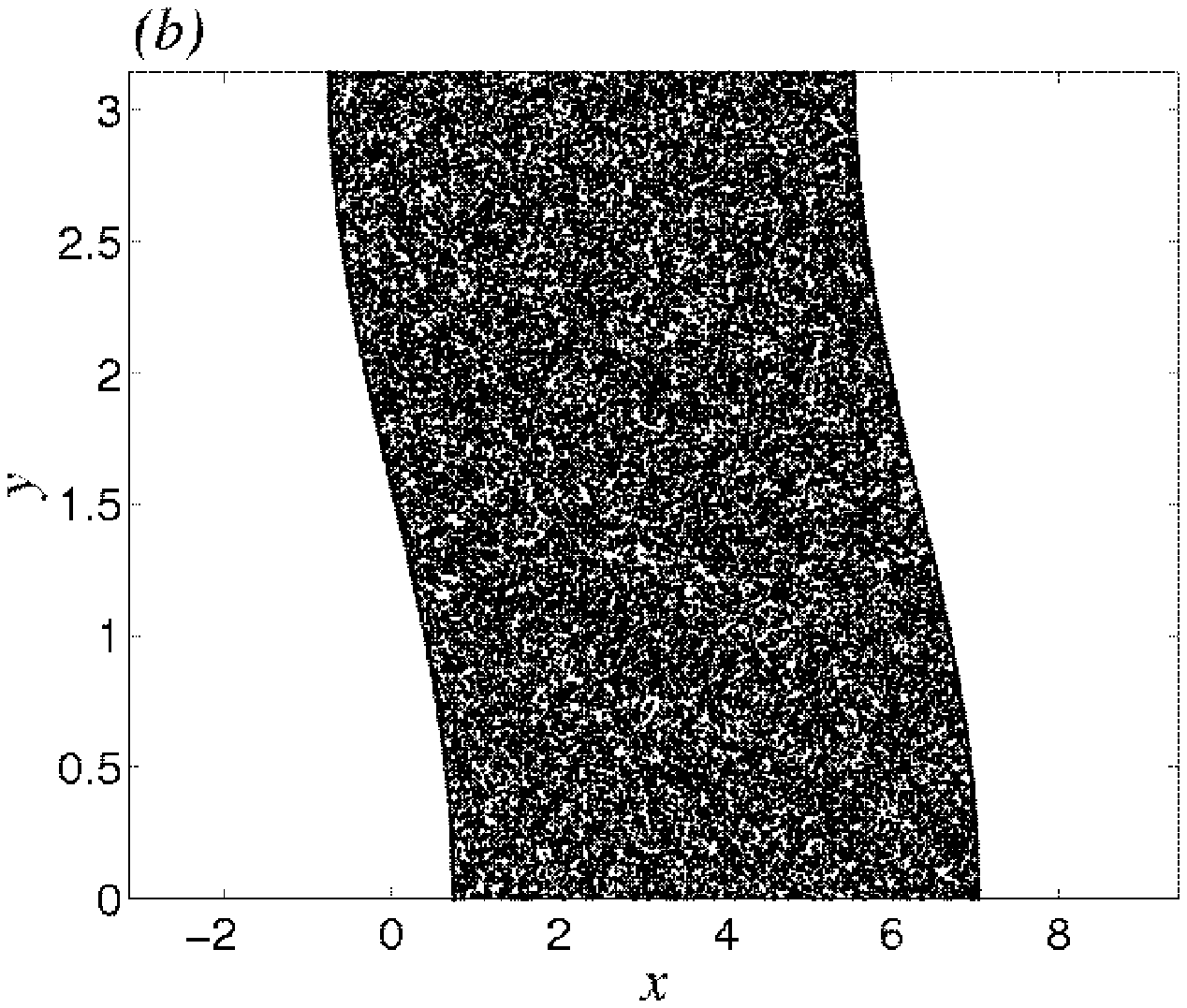}
\par\end{centering}

\caption{\label{Figmix} $(a)$ Streamlines at $t=0$ and $(b)$ Poincar\'{e}
section of the stream function~(\ref{Strmpar}). The parameters are
$\omega=1$ and $\epsilon=0.8$.}
\end{figure}

Figure~\ref{Figmix}$(a)$ depicts the streamlines of the stream
function \begin{equation}
\Psi_{c}(x,y,t)=\sin[x+\epsilon\sin\omega t+\omega^{-1}\cos yC_{\epsilon}(\omega t)]\sin y,\label{Strmpar}\end{equation}
 at $t=0$ for $\omega=1$ and $\epsilon=0.8$. We notice that the
displacement of the rolls remains periodic and parallel to the $x$-direction
with an additional oscillating shear. The Poincar\'{e} section depicted
on Fig.~\ref{Figmix}$(b)$ reveals the dynamics of tracers which
is very different from the one given by the stream function $\Psi_{1}$
given by Eq.~(\ref{courant}) (see Fig.~\ref{Fignmix}$(b)$). The
main differences are that barriers suppressing long range chaotic
transport along the channel are restored around $x=0$ (mod $2\pi$)
(bold curves), and that efficient mixing is achieved within the cell
confined by two barriers. We observe that passive particles appear
to invade the whole confined cell until the fluid is apparently fully
mixed; most of the regular trajectories observed for the stream function
$\Psi_{1}$ are broken by the perturbation. %

\begin{figure}
\begin{centering}
\includegraphics[width=6cm]{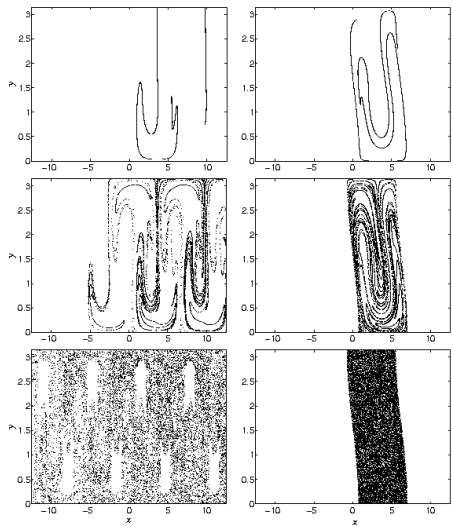}
\par\end{centering}

\caption{\label{Figdrop} Numerical simulation of the dynamics of a dye at
$t=4T$, $t=6T$, $t=18T$ and $T=2\pi$ (from top to bottom)~: left
column for the stream function~(\ref{courant}) and right column
for the stream function~(\ref{Strmpar}). The parameters are $\omega=1$
and $\epsilon=0.8$.}
\end{figure}

In Fig.~\ref{Figdrop}, a numerical simulation of the dynamics of
a dye of tracers in the fluid is shown. The left column depicts the
dynamics of tracers for the stream function~(\ref{courant}). The
right column shows the mixing of a dye within a cell delimited by
two barriers created by the stream function~(\ref{Strmpar}). We
see that the scattering of the dye, which leads to mixing, occurs
through a combination of stretching and folding of the dye in both
cases.

\subsection{Mixing analysis: Local Lyapunov exponent and mean recurrence time
analysis}

As the absence of the regular islands from the Poincar\'{e}
sections does not guarantee the homogeneity of mixing, the study of
local properties of mixing in phase space may provide extensive information.
For this purpose we consider two different types of analysis, namely
the Lyapunov map and the mean recurrence time. Both analysis are performed
within the space of initial conditions.

First, in order to get insight into the action of the perturbation
on the local stability properties of the system, we compute the Lyapunov
map. This method provides local information in phase space. It has
been introduced to detect ordered and chaotic trajectories in the
set of initial conditions. It associates a finite-time Lyapunov exponent
$\nu$ with an initial condition $(x_{0},y_{0})$ at $t=0$.
Let us consider the tangent flow (\ref{tgt_flow}), and define the
maximum finite-time Lyapunov exponent by integrating the flow and
the tangent flow over some time $\tau$ starting with some initial
condition $(x_{0},y_0)$: \[
\nu(x_{0},y_0,\tau)=\frac{1}{\tau}\log\vert\lambda_{\mbox{max}}(x_0,y_0,\tau)\vert,\]
 where $\lambda_{\mbox{max}}(x_0,y_0,\tau)$ is the largest (in norm) eigenvalue
of $J^{\tau}$ (one can also use the eigenvalue of $J^{\tau*}J^{\tau}$
where $J^{\tau*}$ is the transposed matrix of $J^{\tau}$).

From the inspection of the map $(x_{0},y_0)\mapsto\nu(x_{0},y_0,\tau)$ for some given time $\tau$,
one distinguishes the set of initial conditions leading to regular
motion associated with a small finite-time Lyapunov exponent, from
the chaotic ones with larger finite-time Lyapunov exponents. Hence,
this map reveals the phase space structures where the motion of tracers
is trapped on invariant tori, i.e. they highlight islands of stability
located around elliptic periodic orbits. Mixing regions are characterized
by high values of their finite-time Lyapunov exponents.

\begin{figure}
\begin{centering}\includegraphics[width=7cm,keepaspectratio]{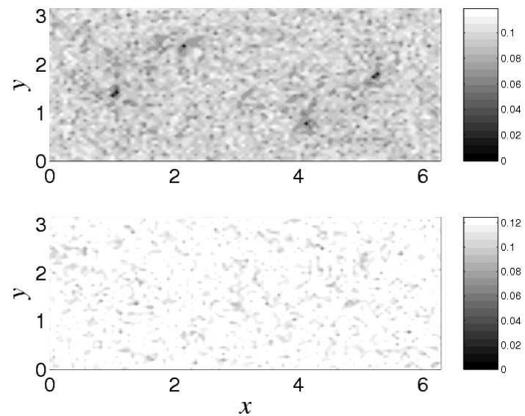} \par\end{centering}

\caption{\label{fig:lce} Lyapunov maps for the dynamics of tracers given
by the stream function $\Psi_{c}$ given by Eq.~(\ref{Strmpar})
for $\omega=1.67$ and $\epsilon=0.63$ (upper panel) and for $\omega=1$
and $\epsilon=0.8$ (lower panel) at a time $T=200\pi$.}
\end{figure}

Figure~\ref{fig:lce} represents the Lyapunov maps for the dynamics
of tracers given by the stream function~(\ref{Strmpar}) for $\omega=1.67$
and $\epsilon=0.63$ (upper panel) and for $\omega=1$ and $\epsilon=0.8$
at a time $T=200\pi$. The dark regions are characteristic of small
values of the Lyapunov exponent. We notice that Fig.~\ref{fig:lce}
(upper panel) shows small remaining islands which are barely noticeable
in the Poincar\'{e} section (see Fig.~3 of Ref.~\cite{Benzekri06}).
For mixing studies, the Lyapunov diagnostic seems to be an appropriate
tool to reveal small non-mixing regions. These regular regions have
disappeared for $\omega=1$ and $\epsilon=0.8$ (lower panel). However,
the Lyapunov map is not able to identify the transport barriers which
means that locally near the barriers, the motion is as chaotic as
inside the cell.

\begin{figure}
\begin{centering}\includegraphics[width=7cm,height=3cm]{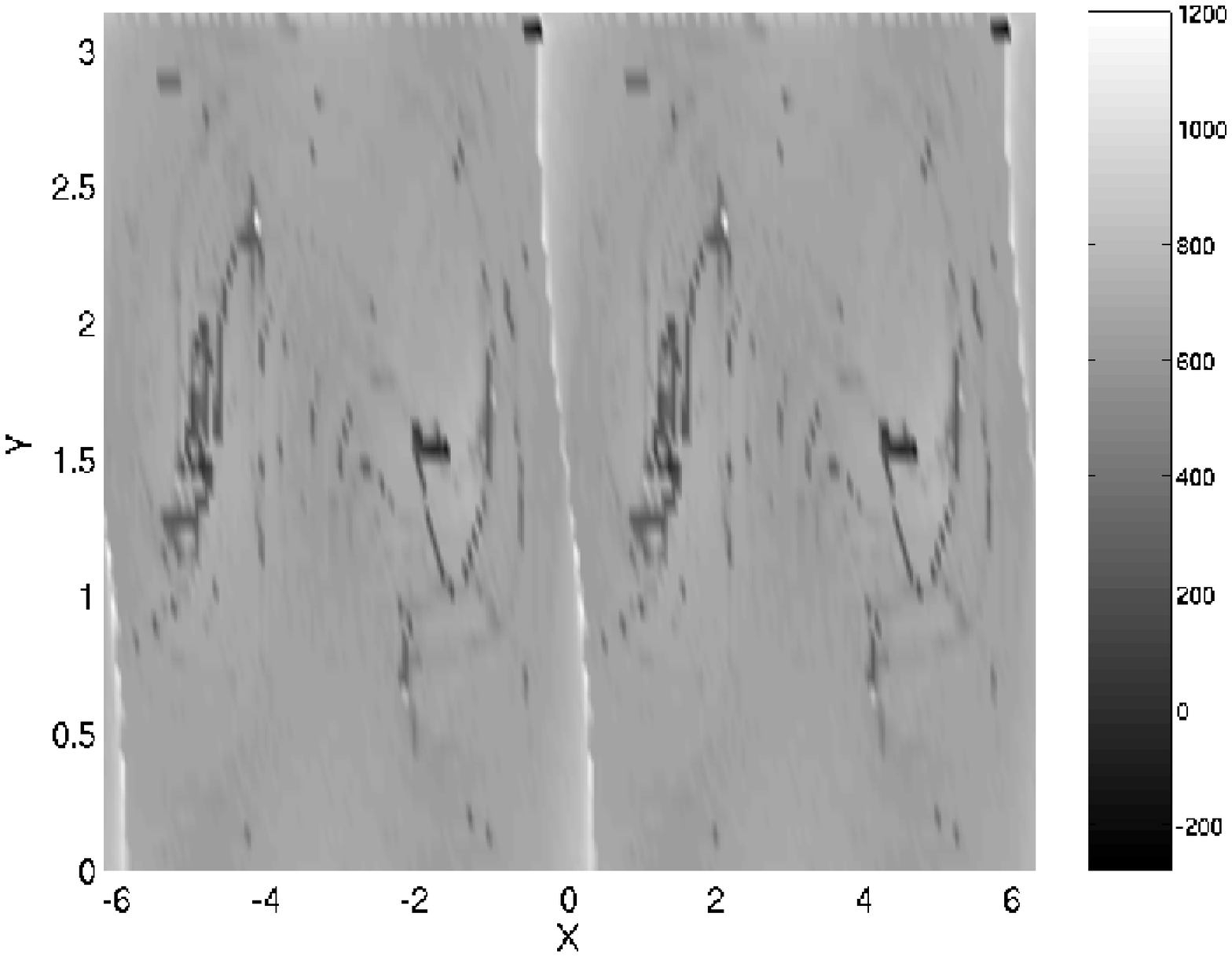}\\
\includegraphics[width=7cm,height=3cm]{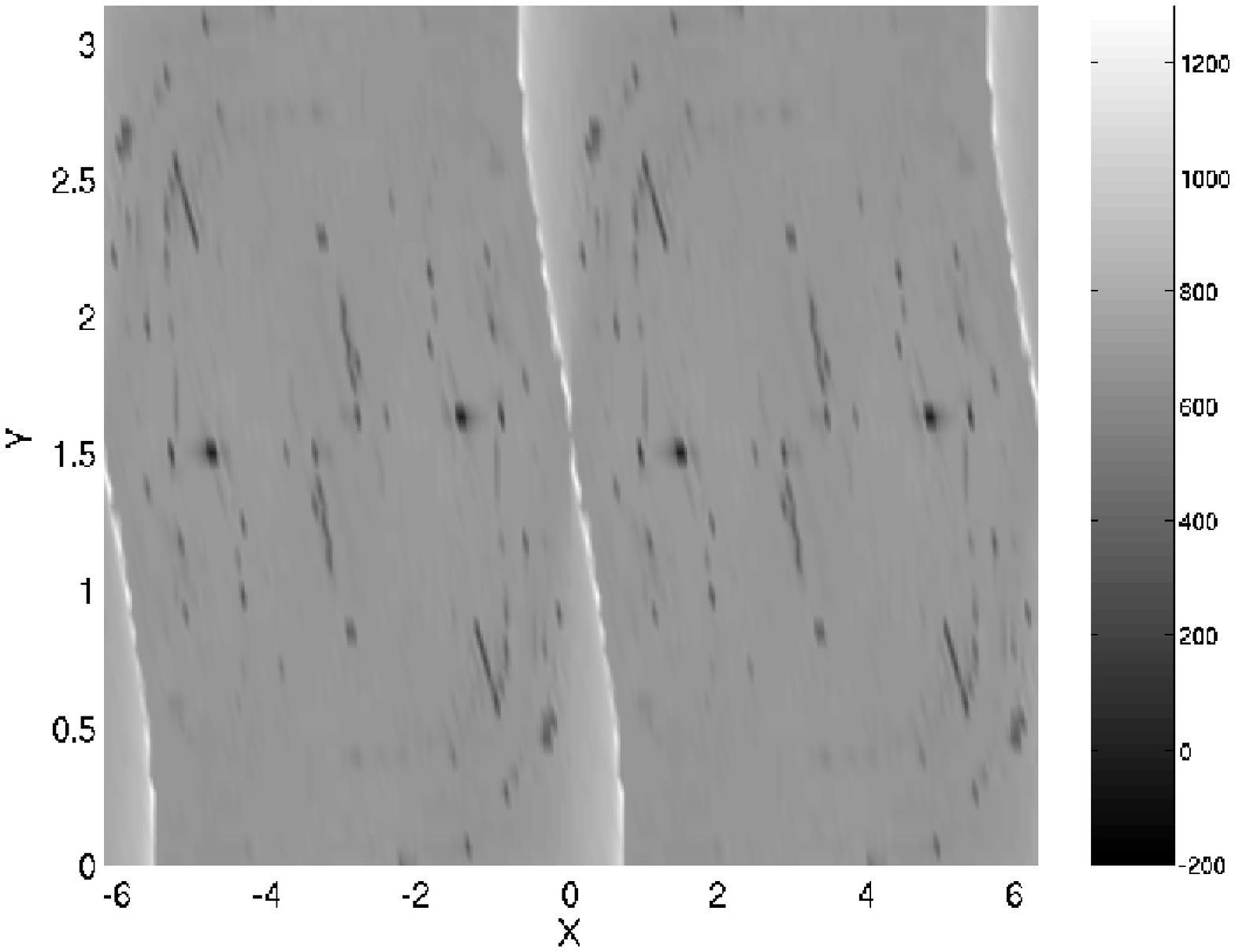}\par\end{centering}

\caption{Mean recurrence time in initial condition space of tracers given
by the stream function $\Psi_{c}$ given by Eq.~(\ref{Strmpar})
for $\omega=1.67$ and $\epsilon=0.63$ (upper panel) and for $\omega=1$
and $\epsilon=0.8$ (lower panel). Initial conditions
space is divided into $100\times 50$ cells. Trajectories are computed for
$\sim10^{6}$ periods, time steps is $dt= T/200$. .\label{fig:Mean-recurrence-time}}
\end{figure}

As seen above, by looking at the Lyapunov map, one can infer local
mixing properties of the flow. However one can notice that since the
created barrier is a separatrix and not a KAM torus as for instance
in Ref.~\cite{Chandre05}, the existence of the barrier cannot be
detected by the Lyapunov map. To complement this analysis, we consider a second diagnostic namely a recurrence
time analysis. An interesting property of return time distributions stems from the
fact that they are known to be sensitive both to local and global
dynamical properties of phase space. For instance, being in the neighborhood
of a hyperbolic periodic orbit versus an elliptic one should affect
the distribution \cite{Haydn05}. Therefore the distribution should
be affected if computed in the neighborhood of a separatrix, or if trapped within a regular region. Since we have already
analyzed the local stability properties of the flow by computing the
Lyapunov map, we will only consider the average first return
time and define a scalar field $\tau(x,y)$ in the space of
initial conditions.
One can indeed assume that if near a given point $(x_{0},y_{0})$
with positive $\nu(x_{0},y_{0})$ the average return time $\tau(x_{0},y_{0})$
is large, then a dye of fluid would explore a large part of phase space and so it would be best to drop the dye in its neighborhood, than
in an other point with similar $\nu$ but a smaller $\tau$.

Typically when analyzing return time statistics from a numerical perspective,
one defines a small area $\Gamma$ in phase space and compute a distribution
of return times of trajectories leaving the area. On one side due
to the fact that, for Hamiltonian systems with finite phase space,
the average return time is finite and scales as $1/\Gamma$ (Kac's
lemma), one has to be cautious not to take $\Gamma$ too small in
order to carry long enough simulations and capture enough events to
build a characteristic distribution. On the other hand, return time
distributions are supposed to be computed for $\Gamma\rightarrow0$.
For the considered flow, due to symmetries we consider $[0,\pi]\times[0,\pi]$
as the space of initial conditions which has been divided regularly
in $2500$ small squares. For each square, the mean return time has
been computed using two trajectories computed for $10^{5}$ periods.
Given the size of $\Gamma$, we collect for each cell about $300$
events. The results are presented in logarithmic scale in Fig.~\ref{fig:Mean-recurrence-time}.
The parameters have been chosen identical to the ones used in Fig.~\ref{fig:lce}.
In contrast to the Lyapunov map, one sees that this diagnostics finds
the barriers (region with long return times). For $\omega=1.67$ and
$\epsilon=0.63$ (upper panel), one can also see small regions with
quite low return times~: They correspond to small regular
islands as mentioned previously. The small return time regions have disappeared
for $\omega=1$ and $\epsilon=0.8$ (lower panel).

Hence, in order to get an accurate picture of the mixing properties of the cell, one has to combine the information of both the local Lyapunov exponent and the local average return time, for example by computing the scalar field $\nu(x_{0},y_{0})\times\tau(x_{0},y_{0})$. Indeed this map shall give us information on good mixing regions,
and provide as well the location of where to drop initially the dye to achieve a faster homogenization and mixing.

\subsection{Robustness}

We have identified a family of perturbations which, while keeping the
cellular structure of the flow, enhance considerably mixing properties
within the cells. The perturbations have been constructed for a very
specific flow; hence a natural question arises in practical situations:
whether or not the considered perturbed system is robust with respect to small changes or errors
in the applied perturbation. Indeed robustness is a key concern in
order for an experimental set up using this type of perturbations.
Below we analyze robustness with respect to four factors: the truncation
of the time series giving the time dependence of the perturbation,
the slip boundary conditions, three dimensional effects, and molecular
diffusivity.

\subsubsection{Truncation of the series}

\begin{figure}
\begin{centering}\includegraphics[width=5cm]{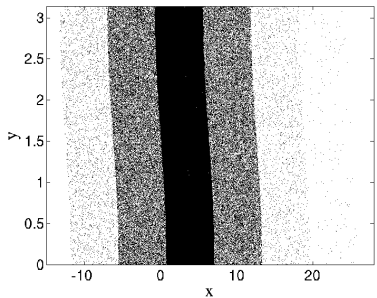} \par\end{centering}
\caption{\label{Figceps} Functions $C_{\epsilon}(t)$ (solid line) and $-2\mathcal{J}_{1}(\epsilon)\cos t$
(dotted line) for $t\in[0,2\pi]$ and $\epsilon=0.8$.}
\end{figure}

From the experimental perspective we may expect some difficulties
in implementing the whole series $C_{\epsilon}(t)$ given by Eq.~(\ref{serbessl}).
One may wonder how the barrier and mixing properties are affected
when one truncates the series and retain for instance only the first
term of the perturbation term (\ref{serbessl}), meaning only one mode is used for the perturbation, $C_{\epsilon}$
is replaced by $-2\mathcal{J}_{1}(\epsilon)\cos t$. Figure~\ref{Figceps}
represents the plot of these two functions for $\omega=1$ and $\epsilon=0.8$.
The small discrepancy between both functions might affect significantly
the transport barriers since it is well known that heteroclinic orbits
are very sensitive to perturbations and are generically broken by
an arbitrarily small perturbation.

\begin{figure}
\begin{centering}\includegraphics[width=5cm]{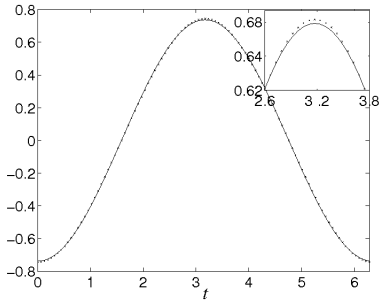} \par\end{centering}
\caption{\label{Fign_truncation} Poincar\'e section computed with the trajectories of 1000 passive tracers
and dynamics given by the stream function~(\ref{eqn:psica}). The
parameters are $\omega=1$, $\epsilon=0.8$. The integration time
is 1000 periods.}
\end{figure}

Trajectories of passive particles with a dynamics given by the stream function
\begin{equation}
\Psi_{c}^{(a)}(x,y)=\sin[x+\epsilon\sin\omega t-2\omega^{-1}\mathcal{J}_{1}(\epsilon)\cos\omega t\cos y]\sin y,\label{eqn:psica}\end{equation}
are displayed in Fig.~\ref{Fign_truncation}. One can see that the
barrier is leaking while mixing properties do not seem to be affected significantly.
One has to notice that given the time length and the amount
of passive particles considered, the leak is small. The
truncated series is still achieving a good {}``targeted mixing''.%

\subsubsection{No slip boundary conditions}

If one wants to take into account the thin boundary layers present
on the boundaries of the channel, it amounts to consider that the
fluid is confined between the two surfaces $y=0$ and $y=\pi$ with no-slip boundary conditions. In this case, the stream function
is modified into~: \begin{equation}
\Psi_0(x,y)=\sin x\; W(y)\:,\label{noslip_courant}\end{equation}
 where no-slip boundary conditions $\dot{x}=\dot{y}=0$ at $y=0$
and $y=\pi$ are obtained with\begin{eqnarray}
W(y) & = & \cos(q_{0}\bar{y})-A_{1}\cosh(q_{1}\bar{y})\cos(q_{2}\bar{y})\nonumber \\
 & + & A_{2}\sinh(q_{1}\bar{y})\sin(q_{2}\bar{y}),\end{eqnarray}
 with $\bar{y}=y/\pi-1/2$, $q_{0}=3.973639$, $q_{1}=5.195214$,
$q_{2}=2.126096$, $A_{1}=0.06151664$ and $A_{2}=0.103887$ (see
Ref.~\cite{chandra}). In this setting the resulting streamlines of the unperturbed stream function~(\ref{noslip_courant}) are very similar to the
ones in Fig.~\ref{Fignmix}$(a)$ with the difference that the velocity
vanishes at the top and bottom of the channel.

\begin{figure}[h!]
\begin{centering}\includegraphics[width=5cm]{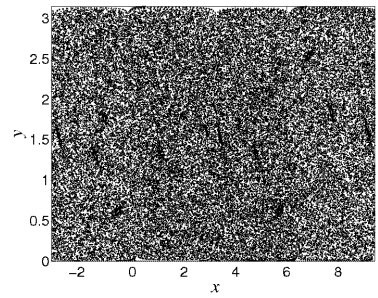} \par\end{centering}
\caption{\label{FigrobSPAD} Poincar\'{e} section of the perturbed stream
function~(\ref{noslip_control}). The parameters are $\omega=1$ and $\epsilon=0.8$. The integration time
is 1000 periods.}
\end{figure}

In order to test the effect of the proposed perturbation~(\ref{control}) on the stream function~(\ref{noslip_courant}), we consider the following perturbed stream function~:
\begin{equation}
\Psi_1^{c}(x,y,t)=\sin[x+\epsilon\sin\omega t+\omega^{-1}\cos yC_{\epsilon}(\omega t)]\; W(y)\:.\label{noslip_control}\end{equation}
Note that the no-slip boundary conditions are
maintained at $y=0$ and $y=\pi$. A poincar\'e section for the stream function~(\ref{noslip_control}) is displayed in Fig.~\ref{FigrobSPAD} for $\omega=1$
and $\epsilon=0.8$. The
barriers are broken and we observe long range chaotic transport
of passive particles along the channel. However, the mixing properties are maintained in the major part of the channel, except for
some areas where some very small regular regions remain.

\begin{figure}

\begin{centering}\includegraphics[width=5cm]{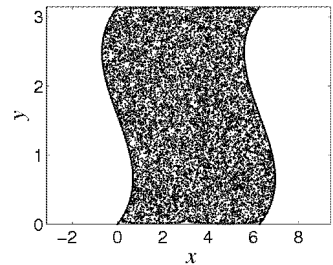} \par\end{centering}

\caption{\label{FigSPAD} Poincar\'{e} section of the stream function~(\ref{contnoslip_courant}). The parameters are
$\omega=1$ and $\epsilon=0.8$.}
\end{figure}

Nevertheless, the exact perturbation of the stream function~(\ref{noslip_courant})
can be derived following the method developed in Sect.~\ref{sec:Constructing barrier}.
It gives the following stream function~: \begin{eqnarray}
\Psi_{c}(x,y,t)=\sin[x+\epsilon\sin\omega t+\omega^{-1}W'(y)C_{\epsilon}(\omega t)]W(y)\nonumber,\\ \label{contnoslip_courant}\end{eqnarray}
which preserves the no-slip boundary conditions at $y=0$ and $y=\pi$.
The resulting Poincar\'{e} section depicted on Fig.~\ref{FigSPAD},
for $\omega=1$ and $\epsilon=0.8$, shows, as in the case of Fig.~\ref{Figmix}(b),
that the stream function~(\ref{contnoslip_courant}) keeps the transport
barriers and the mixing properties. %

\subsubsection{Three dimensional effects}

If the flow is bounded, depending on the size of the boundary layers
and the fluid extension in the third direction, non uniform vorticity
can give rise to a secondary instability leading to a three-dimensional
flow (Ekman pumping). Hence, for the considered perturbed flow, we
may have to take into account the weakly three dimensional case,
which may be given by the empirical flow (see for instance \cite{Solomon_mezic03}):
\begin{eqnarray}
\dot{x} & = & -\sin x\cos y+\epsilon_{1}\sin2x\sin z,\nonumber \\
\dot{y} & = & \cos x\sin y+\epsilon_{1}\sin2y\sin z,\nonumber \\
\dot{z} & = & 2\epsilon_{1}\cos z[\cos2x+\cos2y].\label{eq:Effets_3D}\end{eqnarray}
 Note that the strength of the third component of the flow is characterized
by $\epsilon_{1}$.

\begin{figure}[h!]
\centering \subfigure[$t=10 T$ ]{\label{3DFig$a$}\includegraphics[width=4cm]{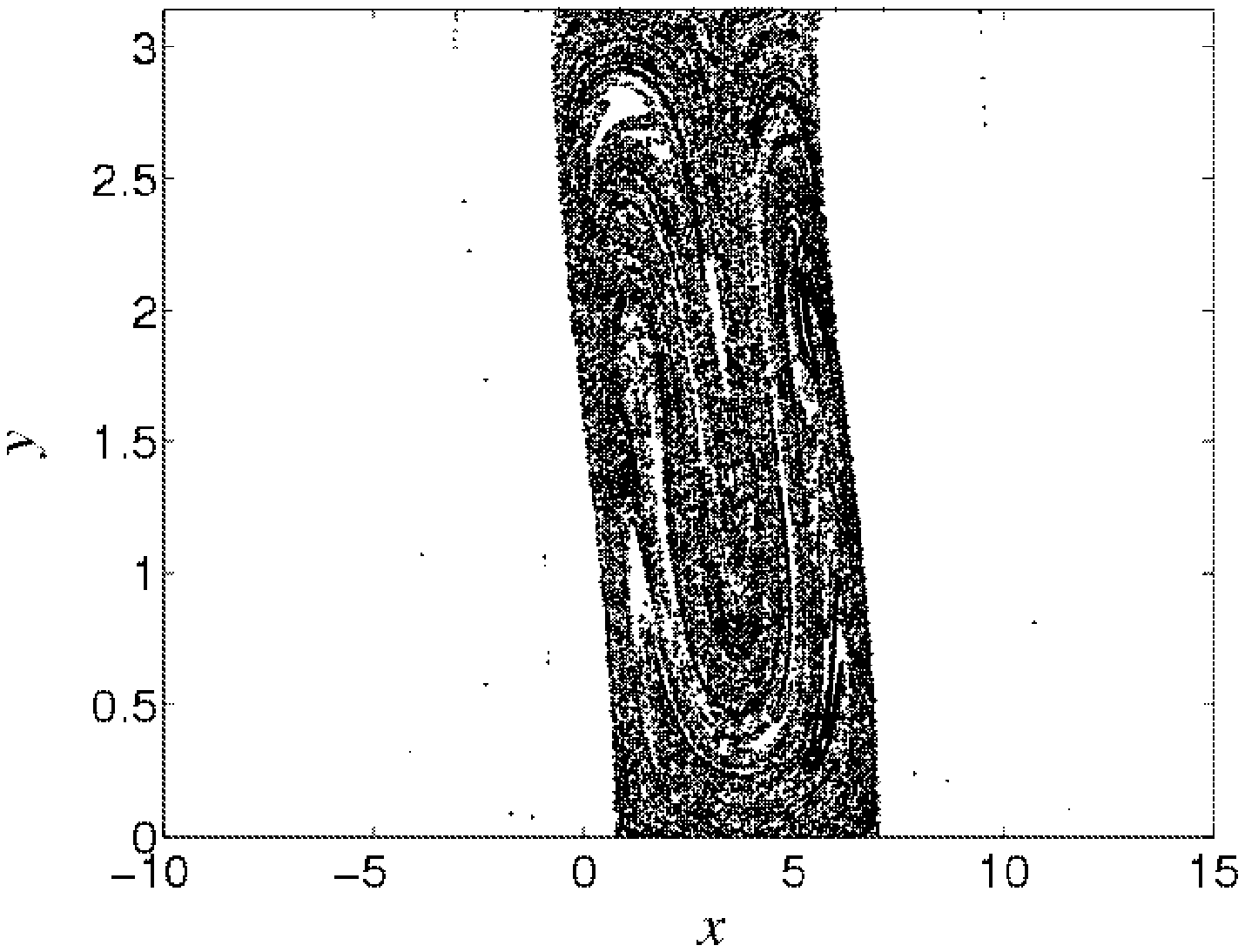}}
\subfigure[$t=100 T$ ]{\label{3DFig$b$}\includegraphics[width=4cm]{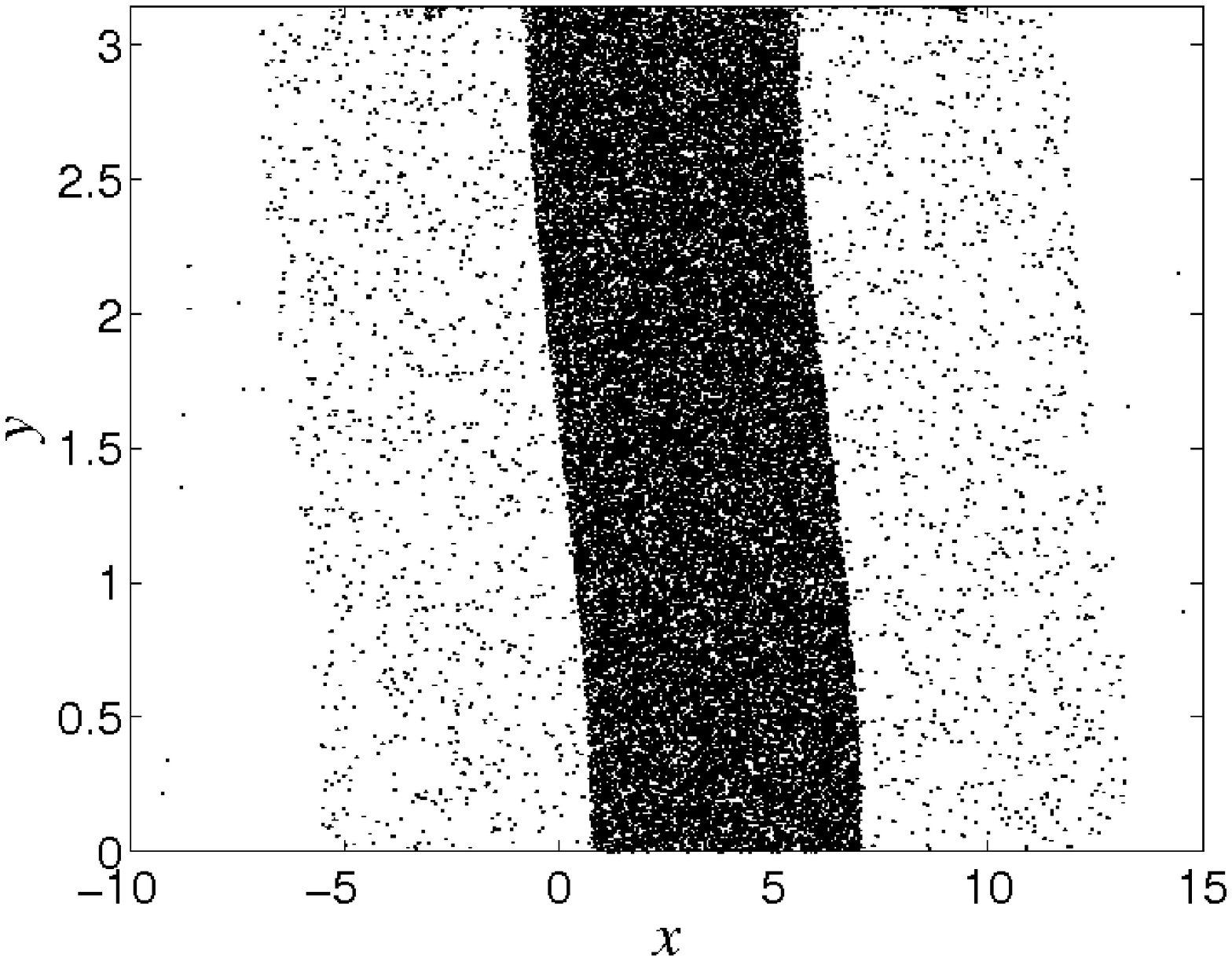}}

\caption{\label{3DFigp} Projection  of trajectories of the system~(\ref{eq:Effets_3D}) for $\omega=1$, $\epsilon=0.8$
and $\epsilon_{1}=0.005$.}
\end{figure}

In order to study how well the two-dimensional barrier fares in this
three-dimensional flow, we apply the perturbation given by Eq.~(\ref{control})
to the right hand side of Eq.~(\ref{eq:Effets_3D}). The perturbed system
is then given by: \begin{eqnarray}
\dot{x} & = & -\sin x_{s}\cos y+\epsilon_{1}\sin2x_{s}\sin z\nonumber \\
 &  & +\omega^{-1}C_{\epsilon}(\omega t)\cos x_{s}\sin^{2}y,\nonumber \\
\dot{y} & = & \cos x_{s}\sin y+\epsilon_{1}\sin2y\sin z,\nonumber \\
\dot{z} & = & 2\epsilon_{1}\cos z[\cos2x_{s}+\cos2y],\label{eq:Effets_3DC}\end{eqnarray}
where $x_{s}=x+\epsilon\sin\omega t+\omega^{-1}\cos yC_{\epsilon}(\omega t)$. We notice that an additional term has been added to $\dot{x}$ in order to ensure a divergence free field.\\
 Keeping a two-dimensional point of view of the system~(\ref{eq:Effets_3DC}),
we visualize the projections in the $(x,y)$ plane of the position
of passive tracers. When considering $\epsilon_{1}=0.005$ and $4.10^{4}$
passive particles at time $t=10T$, where $T=2\pi$, an effective barrier remains as it is shown in Fig.~\ref{3DFig$a$} and mixing properties are
not affected, but we observe some advected particles which escape
from the cell. However as depicted in Fig.~\ref{3DFig$b$}, when
the integration time is $t=100T$ (for the same
value of $\epsilon_{1}$ and number of particles), the barrier still
influences the motion but leaks since a more significant number of particles
get through these barriers.

\subsubsection{Molecular diffusivity}

\begin{figure}
\begin{centering}\includegraphics[width=7cm]{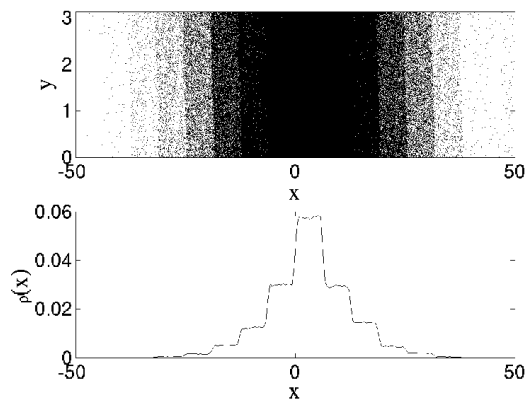}\par\end{centering}

\caption{Trajectories of 1000 passive tracers given by Eq.~(\ref{eq:Langevin}).
(top). Local density of tracers (bottom). The parameters are $\mu=4.10^{-2}$ (Peclet number $P_{e}\approx600$),
$\omega=1$ and $\epsilon=0.8$. The integration
time is 1000 periods. \label{fig:Influence-of-molecular}}
\end{figure}

Finally, we have up to now considered ideal passive tracers, which
are not subject to any molecular diffusivity, which may be a good
approximation for high Peclet numbers. However it is likely that for
any finite molecular diffusivity the barrier will leak. In order to
illustrate this phenomenon, we consider that tracers are actually
subject to a Langevin equation associated with the stream function
$\Psi_{c}$ given by Eq.~(\ref{Strmpar})~:

\begin{equation}
\dot{x}=-\frac{\partial\Psi_{c}}{\partial y}+b_{x}(t),\hspace{1.2cm}\dot{y}=\frac{\partial\Psi_{c}}{\partial x}+b_{y}(t)\:,\label{eq:Langevin}\end{equation}
 where $b_{x}(t)$ and $b_{y}(t)$ are two independent delta-correlated
white noises, with zero mean and a given amplitude $\mu$. Numerical
results are displayed in Fig.~\ref{fig:Influence-of-molecular},
where $\mu=4.10^{-2}$ (which corresponds to Peclet number of $P_{e}\approx600$),
$\omega=1$ and $\epsilon=0.8$. In order to avoid crossing across
the {}``walls'' $y=\pi$ and $y=0$, we took $b_{y}=0$. %

One can see that for this type of Peclet values, the transport barrier
is indeed leaking, and in fact  the truncation of $C_{\epsilon}(t)$ does
not affect how much the barrier leaks.

In summary the robustness of the proposed perturbation in different settings has been investigated. One may infer that the most drastic effects are induced by boundary conditions. On the other hand provided that the right perturbation is computed, one concludes that if three-dimensional effects are weak and Peclet number is high enough, a truncation to the first term of the series $C_{\epsilon}(t)$ is sufficient, and actually implementing more terms seems useless.

\subsection{Targeted mixing regimes}

It was shown in Sec.~\ref{sec:Constructing barrier} that a perturbation
allows one to localize tracers into a finite volume of phase space.
However, depending on the values of the parameters of the perturbation
$(\omega,\epsilon)$ in Eq.~(\ref{Strmpar}), some regular islands may exist and prevent complete
mixing inside the cell. In this section we propose to identify the
domain of parameters (if any) such that these islands do not exist
and the cell become fully mixing.

This domain in parameter space can be determined by analyzing the
linear stability of a few periodic orbits of the system as discussed
in Sec.~\ref{sec:residus}. These orbits are those with low rotation
numbers, around which the resonant islands organize. Indeed, an island is organized around a central elliptic
orbit~: If the latter were to turn hyperbolic, the dynamics might
become (locally) chaotic, since generically chaos is expected in the
neighborhood of hyperbolic orbits by an infinite number of intersections
between its stable and unstable manifolds. In order to monitor these
orbits, we use a scalar indicator of linear stability, such as the
residue (see Sec.~\ref{sec:residus}). The fully mixing regime will
then correspond to the residues of selected periodic orbits being
below $0$ or above $1$. However, the change of linear stability
of a given periodic orbit might not change its nonlinear stability
and still preserve invariant tori in its neighborhood. This linear stability
analysis of periodic orbits has to be completed by an
a posteriori check to determine whether or not the regular island
has disappeared with the elliptic periodic orbit turning hyperbolic.

Let us consider the following range of parameters $(\omega,\epsilon)\in[0.6,2.2]\times[0,2.2]$. Inspection of several Poincar\'e sections reveals that in this range,
it is essentially the nature of eight periodic orbits which drives the
mixing properties inside the cell. However, thanks to the symmetry with
respect to the point $(x,y)=(\pi,\pi/2)$, this set reduces to only
four orbits~: Three of them have rotation number one, namely $\mathcal{O}_{1}$, $\mathcal{O}_{1}^{\alpha}$
and $\mathcal{O}_{1}^{\beta}$, while the fourth one, called $\mathcal{O}_{2}^{+}$
has a rotation number $Q=2$. Depending on the value of the parameters
$(\omega,\epsilon)$, these orbits can be elliptic - and create resonant
islands around them -, or hyperbolic. In order to describe the nature
of each of these four orbits for a given value of parameters, we
use the nomenclature $[N(\mathcal{O}_{1})N(\mathcal{O}_{2}^+)N(\mathcal{O}_{1}^{\alpha})N(\mathcal{O}_{1}^{\beta})]$,
with $N(\mathcal{O})=h$ if $\mathcal{O}$ is hyperbolic, $e$ if
elliptic, and $0$ if it does not exist. Let us precise that there
also exists a hyperbolic orbit with $Q=2$,
called $\mathcal{O}_{2}^{-}$, which forms a Birkhoff pair with $\mathcal{O}_{2}^{+}$.
As it remains hyperbolic in the range of parameters under consideration,
it only provides a better understanding of the bifurcation process,
but does not influence the mixing properties inside the cell.

\begin{figure}[htp]
\centering \subfigure[$\omega=1 , \epsilon=0.8$]{\label{epsom1}\includegraphics[width=4cm]{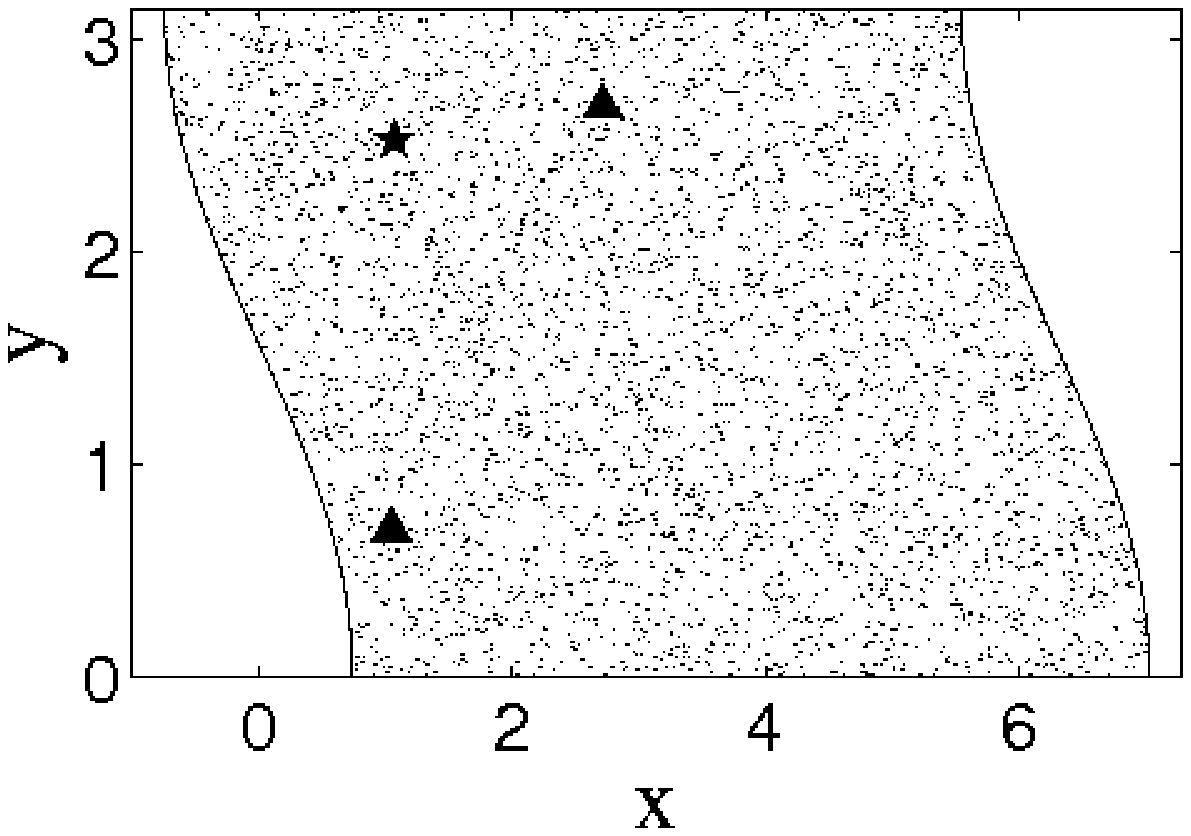}}
\subfigure[$\omega=1.8 , \epsilon=0.5$]{\label{epsom2}\includegraphics[width=4cm]{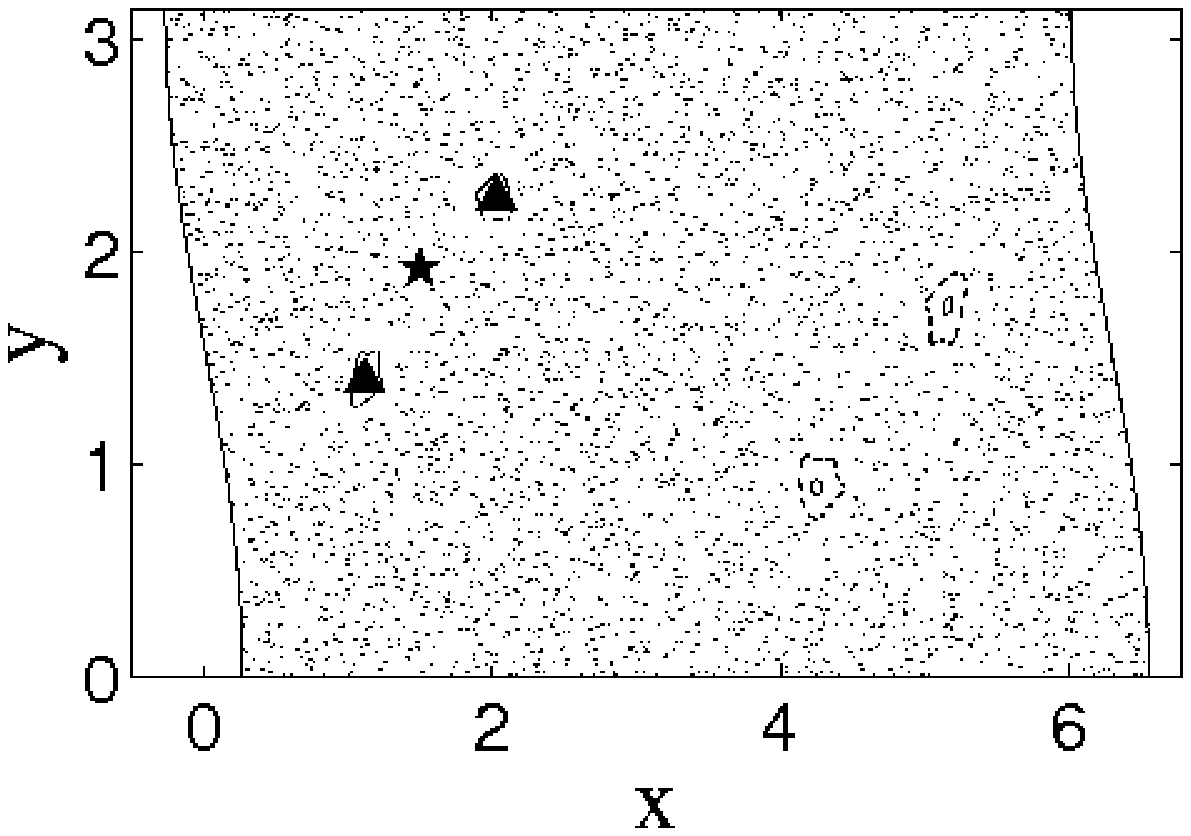}}\\
\subfigure[$\omega=1.2 , \epsilon=0.2$]{\label{epsom3}\includegraphics[width=4cm]{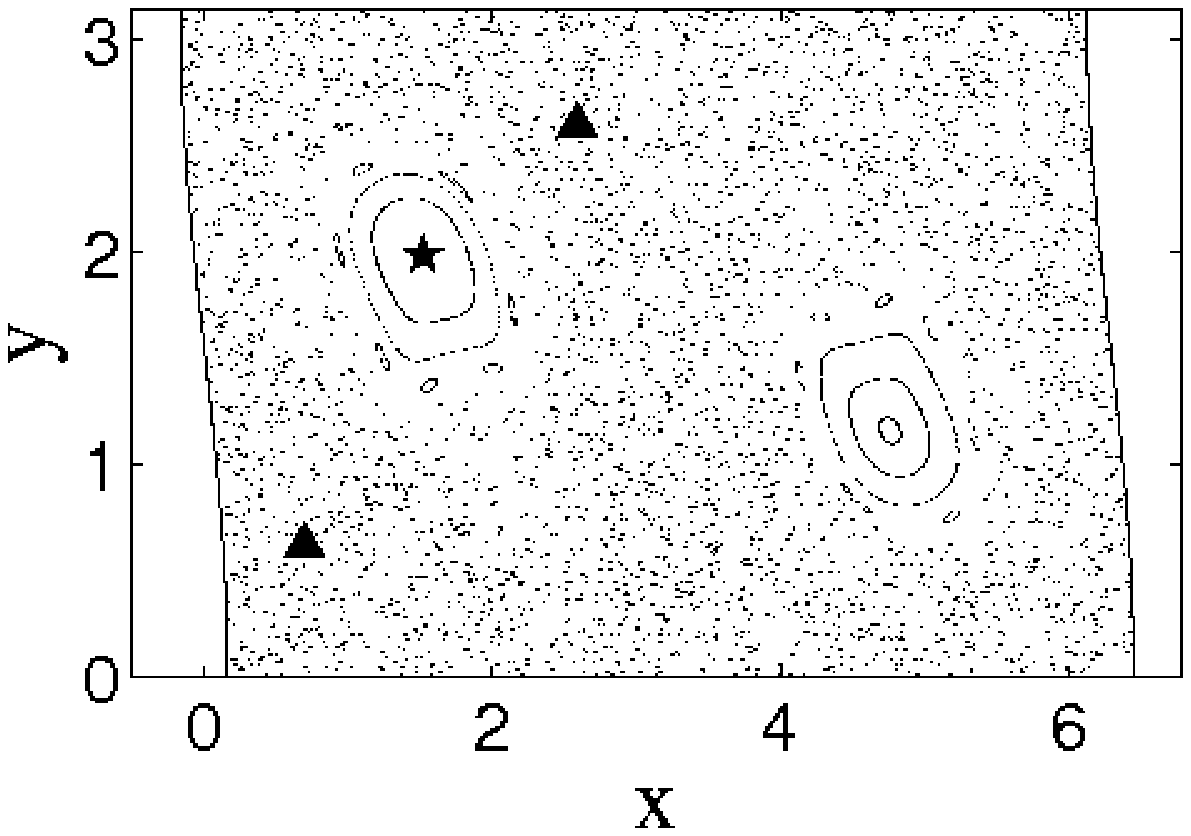}}
\subfigure[$\omega=1.6 , \epsilon=0.2$]{\label{epsom4}\includegraphics[width=4cm]{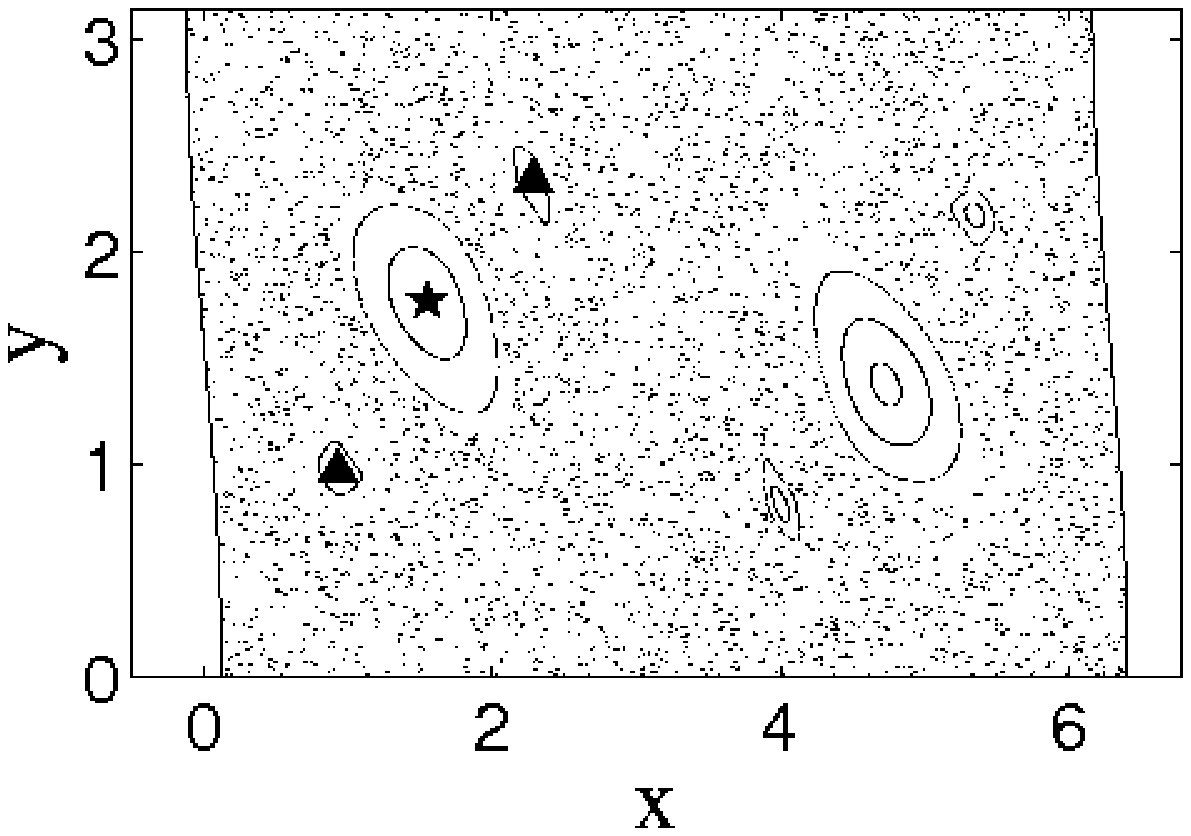}}\\
\subfigure[$\omega=0.58 , \epsilon=1.25$]{\label{epsom5}\includegraphics[width=4cm]{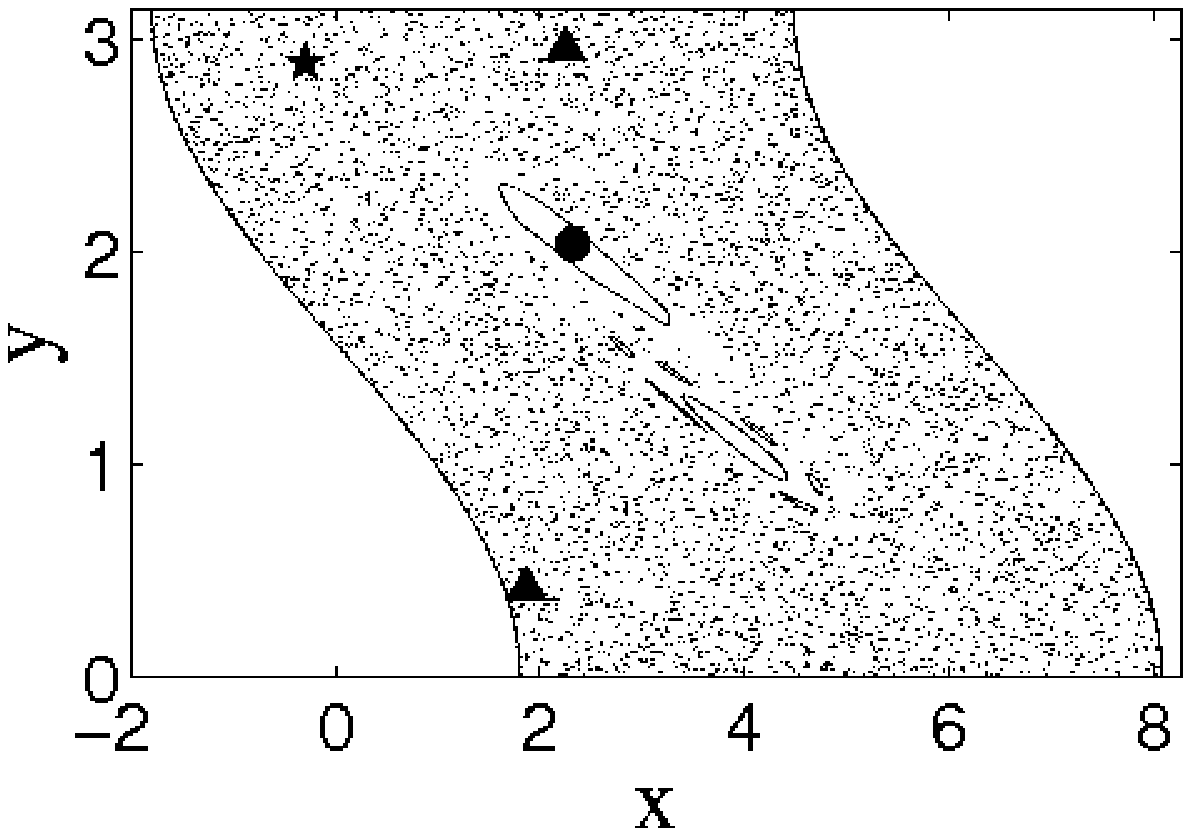}}
\subfigure[$\omega=0.58 , \epsilon=0.6$]{\label{epsom8}\includegraphics[width=4cm]{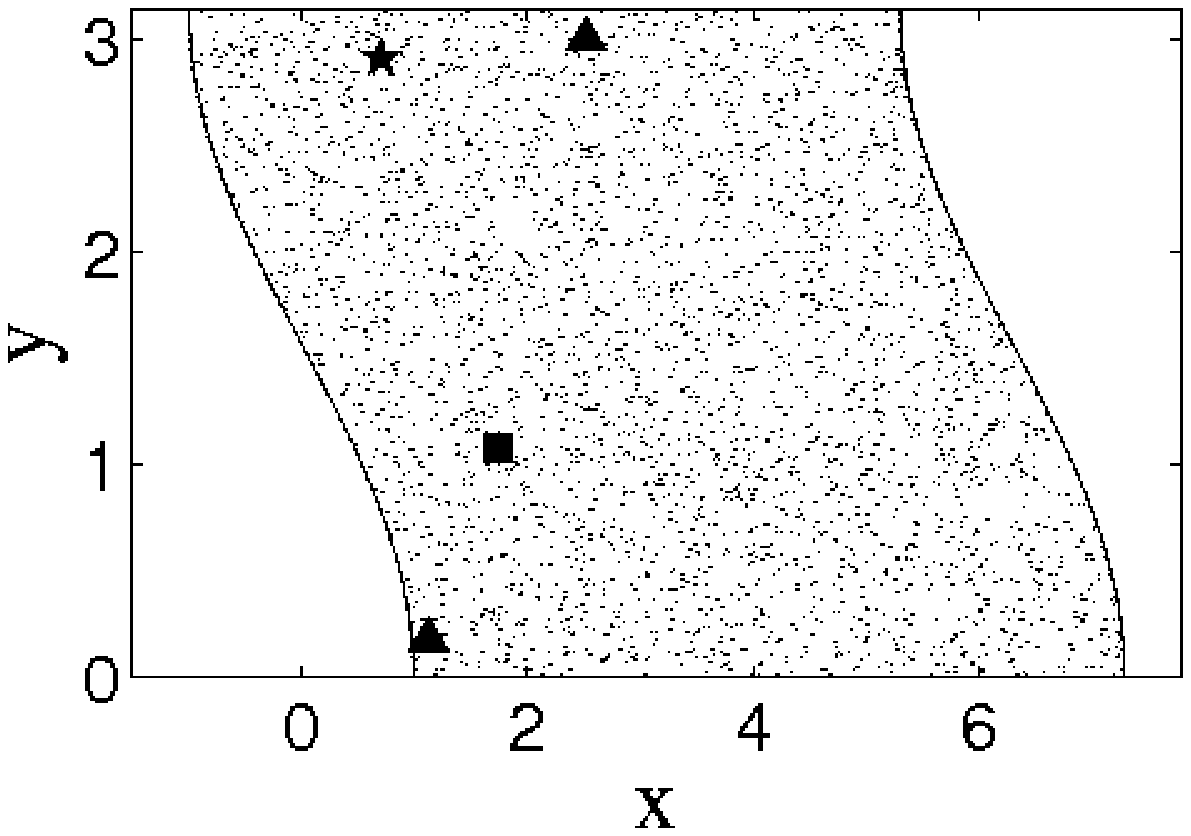}}\\
\subfigure[$\omega=2.1 , \epsilon=0.25$]{\label{epsom7}\includegraphics[width=4cm]{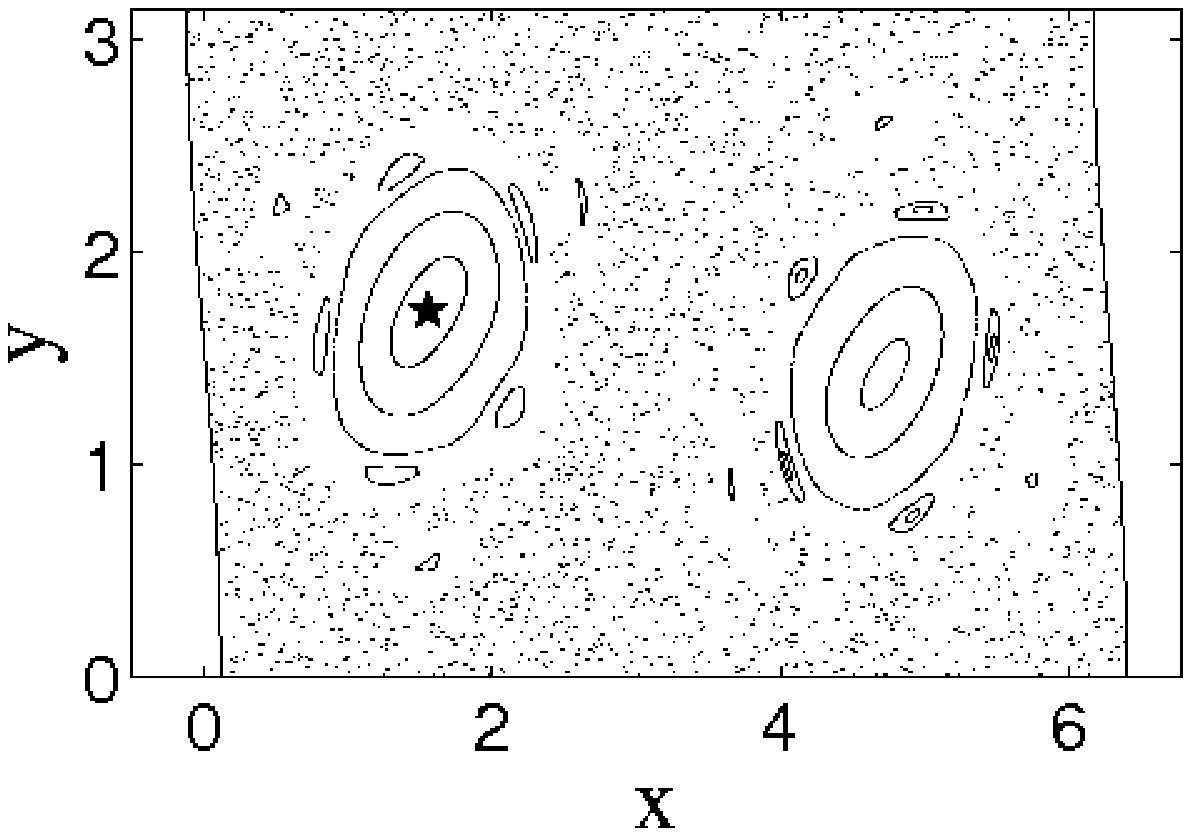}}
\subfigure[$\omega=0.63 , \epsilon=0.01$]{\label{epsom6}\includegraphics[width=4cm]{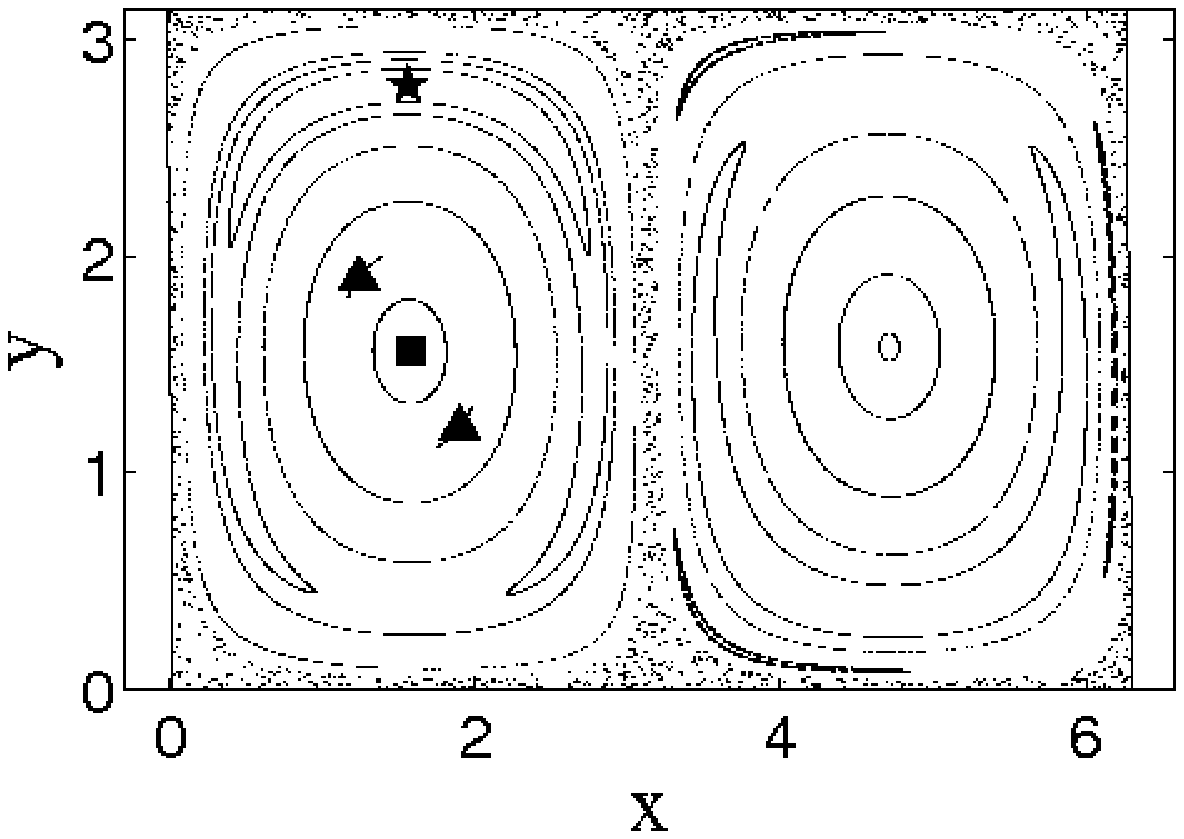}}
\caption{Poincar\'{e} sections in the different domains of parameters $(\omega,\epsilon)$ for the stream function (\ref{Strmpar}).
The stars, full circles, squares and triangles indicate respectively the locations of $\mathcal{O}_{1}$, $\mathcal{O}_{1}^{\alpha}$, $\mathcal{O}_{1}^{\beta}$ and $\mathcal{O}_{2}^{+}$.\label{eightsec}}
\end{figure}

Figure~\ref{eightsec} depicts Poincar\'{e} sections for eight different
values of parameters. These cases illustrate some possible mixing
regimes in the considered range of parameters. Figures~\ref{epsom1}-\ref{epsom4}
(corresponding respectively to $[hh00]$, $[he00]$, $[eh00]$ and $[ee00]$)
show how $\mathcal{O}_{1}$ and $\mathcal{O}_{2}^{+}$ can coexist
in their two forms (elliptic and hyperbolic), the full mixing regime
being reached when both are hyperbolic as in Fig.~\ref{epsom1} where it is
$[hh00]$. We notice that when ${\mathcal O}_1$ or ${\mathcal O}_2^+$ is elliptic, the mixing is only prevented by the elliptic island. Only small secondary islands are observed (for instance, in Fig.~\ref{epsom3}). This reinforces the importance of the considered set of periodic orbits for mixing properties.

For smaller values of $\omega$, the orbits $\mathcal{O}_{1}^{\alpha}$ and $\mathcal{O}_{1}^{\beta}$ may also appear (if $\omega\leq 1.2$ and large $\epsilon$ for $\mathcal{O}_{1}^{\alpha}$, or $\omega\leq 1$ and small $\epsilon$ for $\mathcal{O}_{1}^{\beta}$), and their nature have to
be taken into account, as one can see on Fig.~\ref{epsom5} $[hhe0]$,
where the hyperbolicity of $\mathcal{O}_{1}$ and $\mathcal{O}_{2}^{+}$
is not sufficient to ensure the full mixing inside the cell, as $\mathcal{O}_{1}^{\alpha}$ is present in its elliptic form. Mixing can still be obtained in presence of $\mathcal{O}_{1}^{\beta}$,
as can be seen on Fig.~\ref{epsom8} which corresponds to the case $[hh0h]$. The opposite case $[ee0e]$
is depicted on Fig.~\ref{epsom6} where all three orbits
are elliptic and almost no mixing occurs.

Finally, one can see on Fig.~\ref{epsom7} which is $[e000]$ how for large
values of $\omega$ (typically beyond $2$), $\mathcal{O}_{2}^{+}$
no more exists, while $\mathcal{O}_{1}$ stays elliptic~: Furthermore,
new resonant islands have appeared, associated with new periodic
orbits (with $Q=3$ and $Q=5$ as shown). This illustrates the fact
that in another range of parameters, the dynamics may be guided by
higher order periodic orbits which are associated with smaller islands.

\begin{figure}[htp]
\centering
 \subfigure[$\omega=1.42$]{\label{om142}\includegraphics[width=8cm]{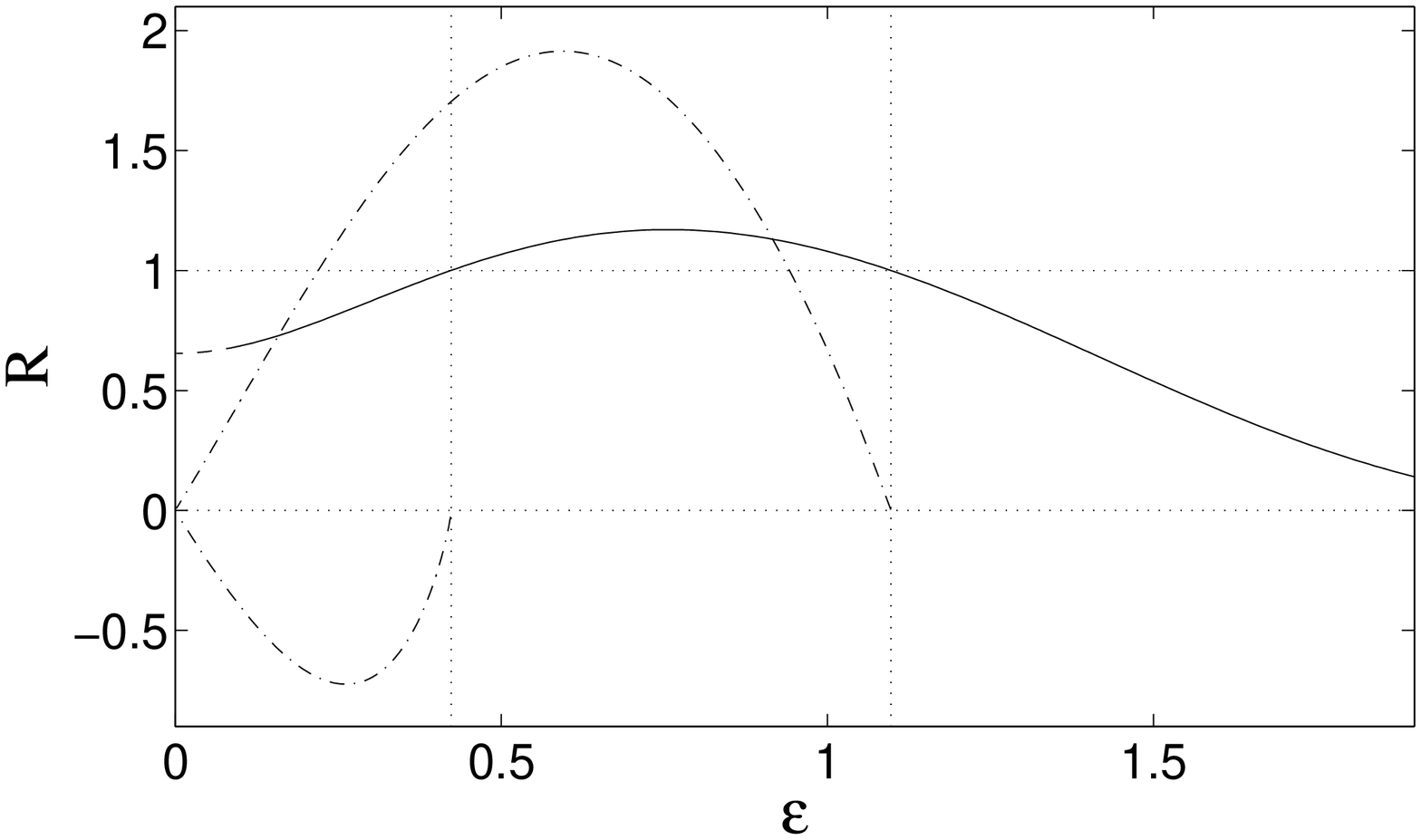}}
\\
 \subfigure[$\omega=1.67$]{\label{om167}\includegraphics[width=8cm]{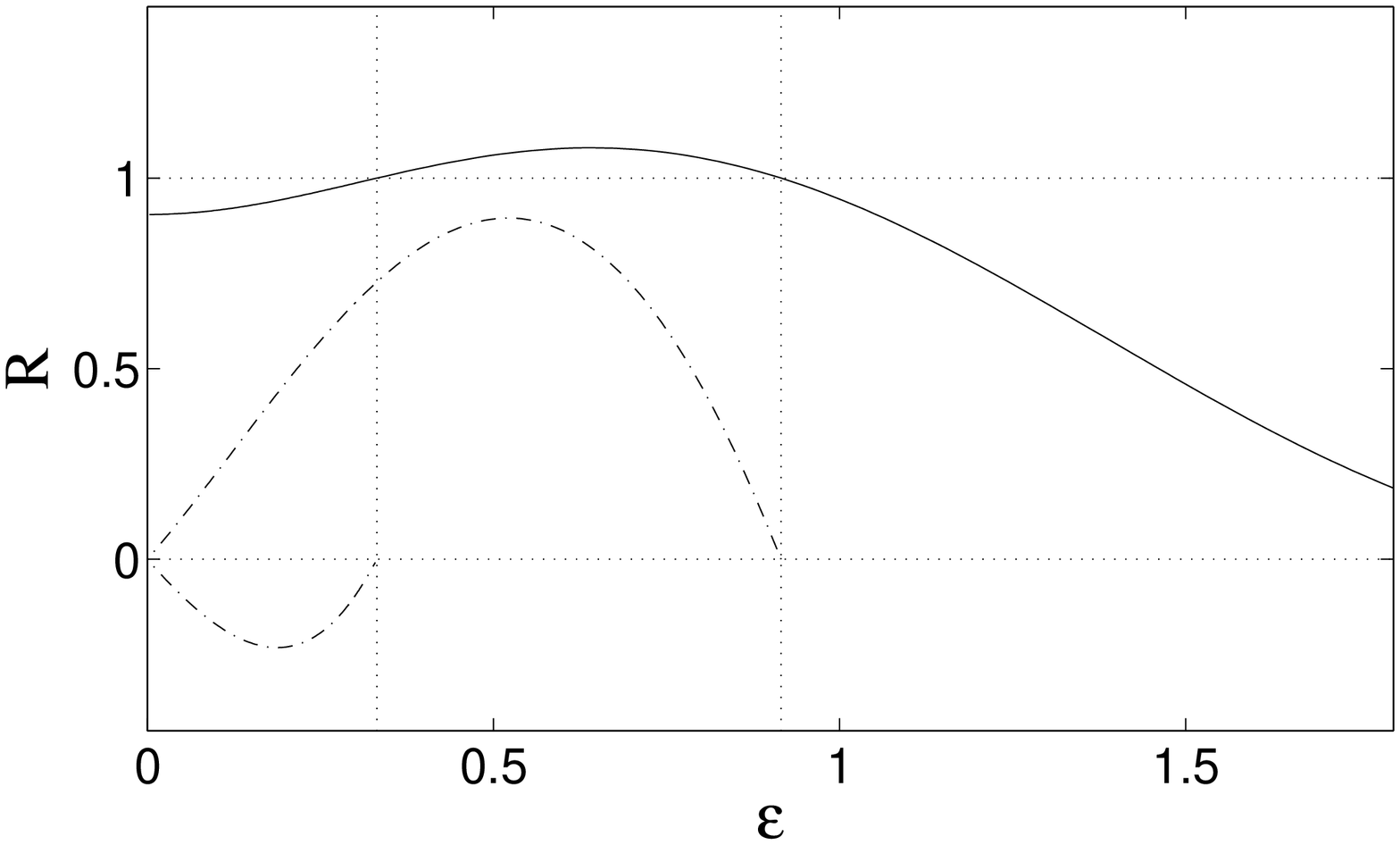}}
\\
 \subfigure[$\omega=0.8$]{\label{om08}\includegraphics[width=8cm]{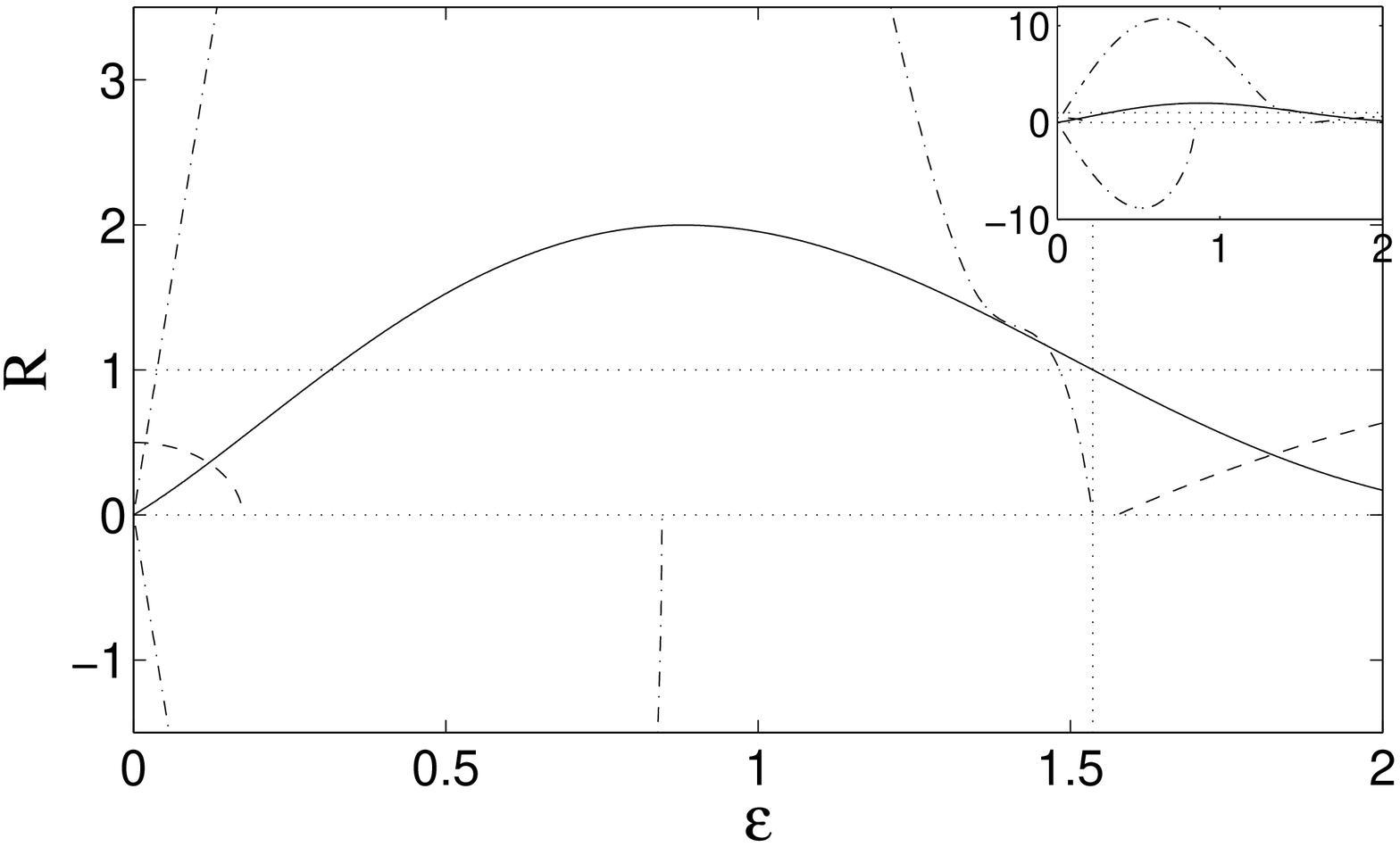}}
\\

 \subfigure[$\omega=0.58$]{\label{om058}\includegraphics[width=8cm]{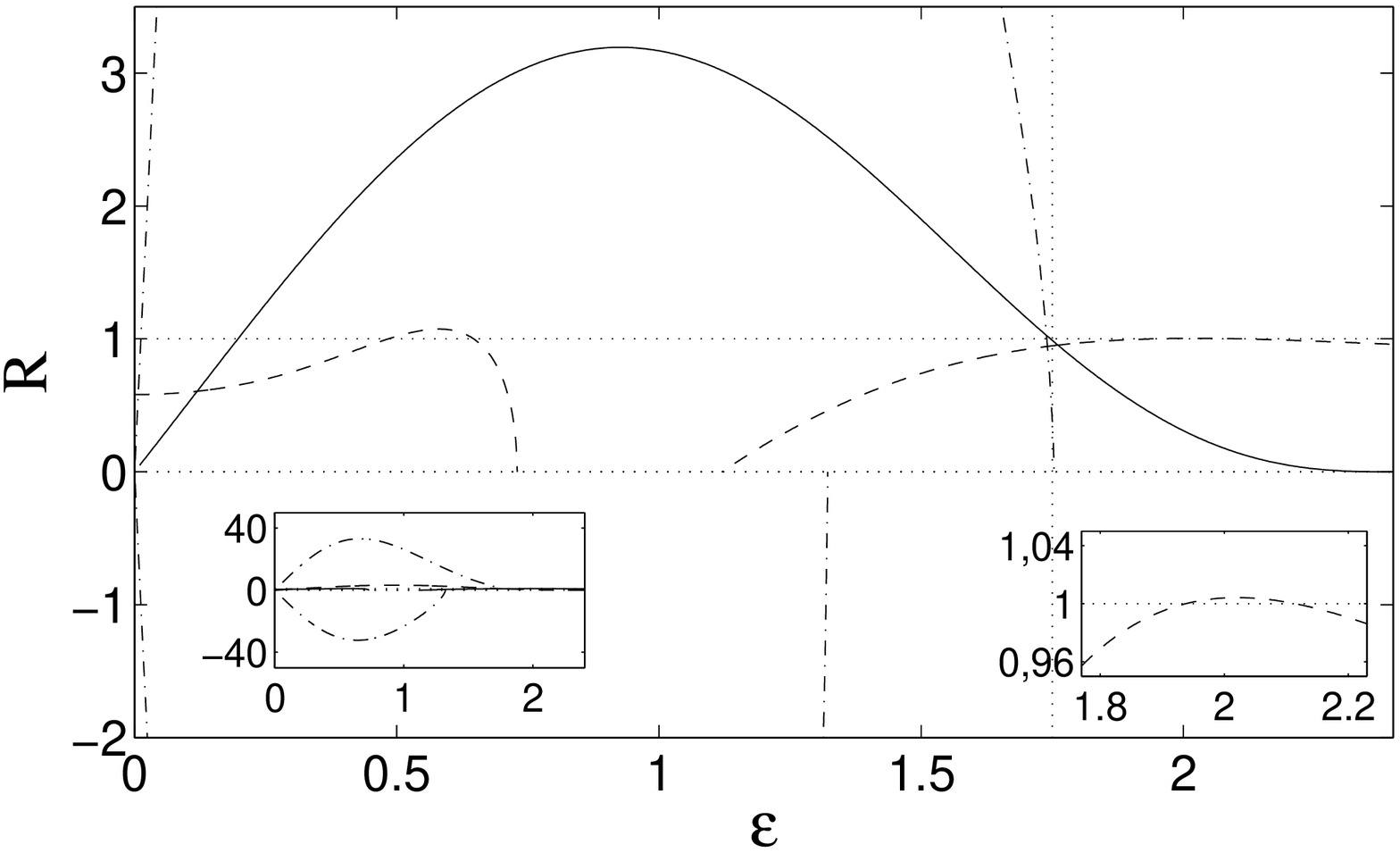}}
\\
 \subfigure[$\omega=2.08$]{\label{om208}\includegraphics[width=8cm]{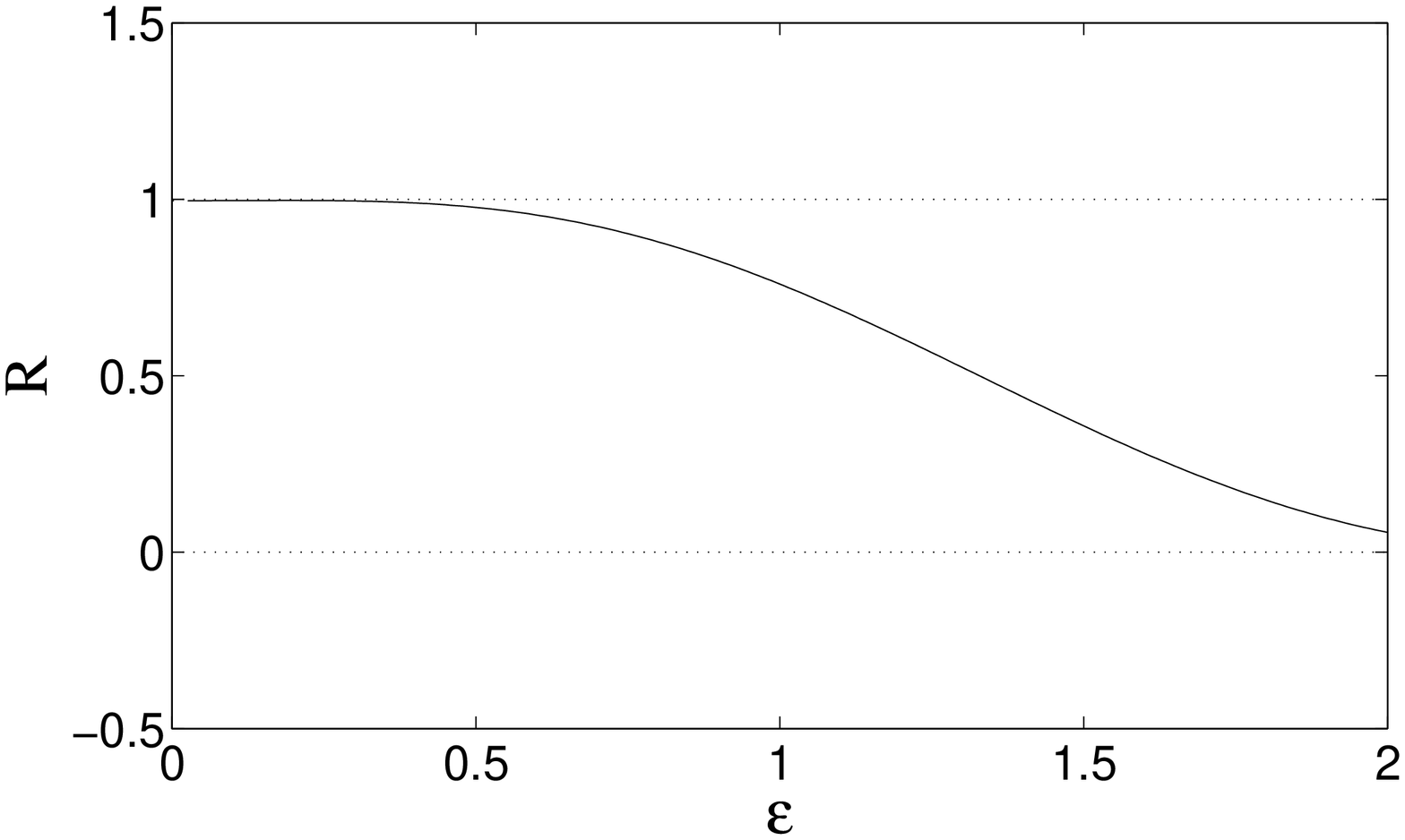}}
\\
\caption{Residue curves for the stream function~(\ref{Strmpar}) as functions of the amplitude $\epsilon$~: The plain line corresponds to the residues of $\mathcal{O}_{1}$,
the dashed one to the residues of $\mathcal{O}_{1}^{\beta}$ and the dash-dotted
to the ones of $\mathcal{O}_{2}^{+}$ (upper curve) and $\mathcal{O}_{2}^{-}$
(lower curve). The dotted line indicates the locations of the bifurcations. \label{omcst}}
\end{figure}

A better insight in the bifurcation scheme can be gained by varying only one parameter at a time. Figure~\ref{omcst} depicts the residue curves associated with the orbits $\mathcal{O}_{1}$, $\mathcal{O}_{2}^{+}$, $\mathcal{O}_{2}^{-}$, $\mathcal{O}_{1}^{\alpha}$ and $\mathcal{O}_{1}^{\beta}$ when varying $\epsilon$ and keeping $\omega$ constant. Two curves have been computed in the intermediate frequency range $\omega=1.42$ and $\omega=1.67$. In the intermediate regime, only three of these five orbits exist, namely $\mathcal{O}_{1}$ (plain line), $\mathcal{O}_{2}^{+}$ and $\mathcal{O}_{2}^{-}$ (resp. upper and lower dash-dotted lines).

For $\omega=1.42$ (see Fig.~\ref{om142}) and for low $\epsilon$, both $\mathcal{O}_{1}$ and $\mathcal{O}_{2}^{+}$ are elliptic~: only partial mixing occurs. While $\mathcal{O}_{2}^{+}$ turns hyperbolic as $\epsilon$ is increased (around $\epsilon\approx 0.2$), $\mathcal{O}_{1}$ remains elliptic until $\epsilon \approx 0.4$. $\mathcal{O}_{1}$ turns from elliptic to hyperbolic by merging with $\mathcal{O}_{2}^{-}$, according to the process~:
\begin{equation}\label{bifheh} \mathcal{O}_{2}^{-}[h] + \mathcal{O}_{1}[e] \rightarrow \mathcal{O}_{1}[h].\end{equation}

\begin{figure}[ht]
\centering \includegraphics[width=7cm]{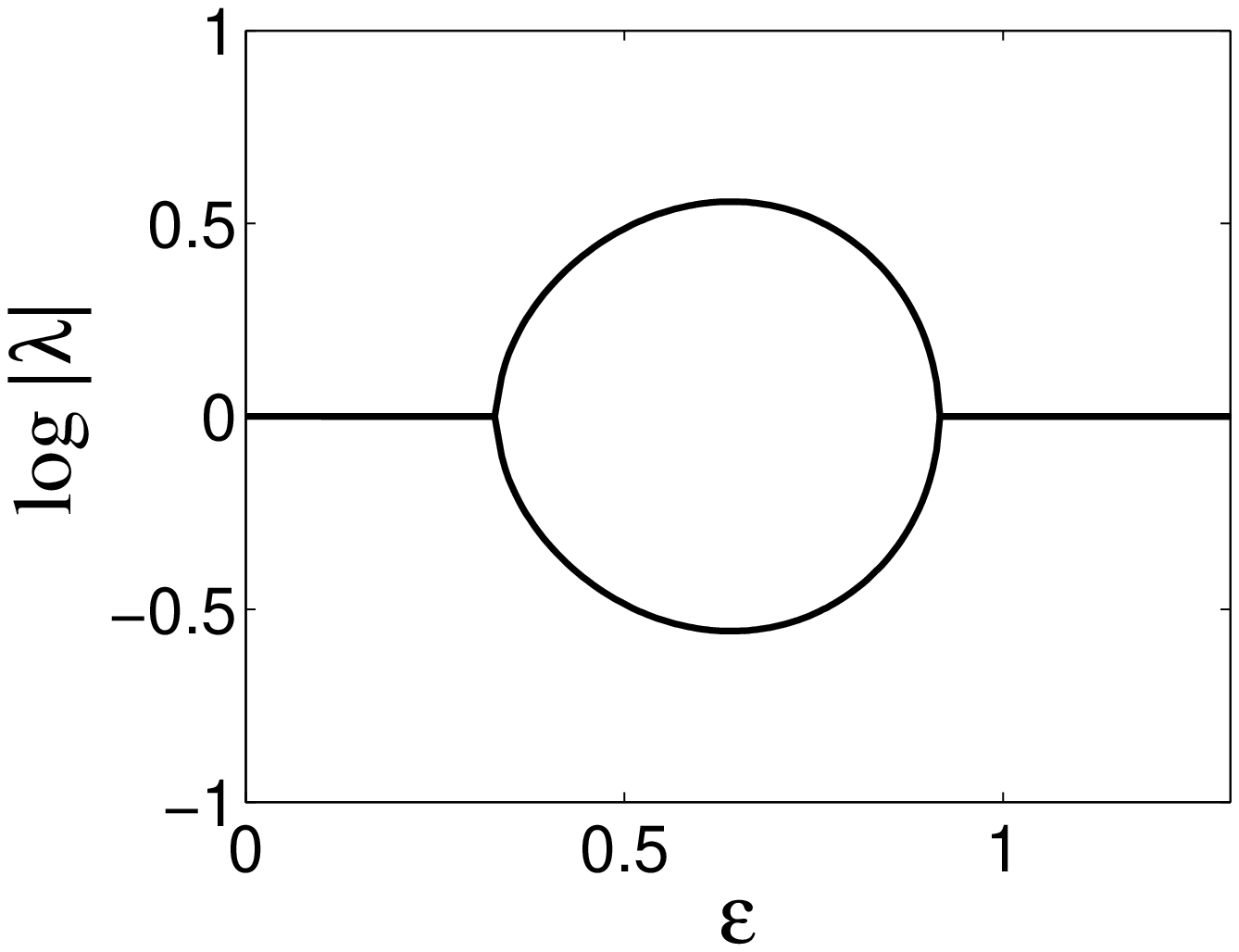}
\caption{Modulus of the eigenvalues of orbit $\mathcal{O}_{1}$ for the stream function~(\ref{Strmpar}) with $\omega=1.67$.\label{bifu}}
\end{figure}

This bifurcation does not happen just for this particular choice of parameters, but appears to be fairly generic in this range of parameters. For a better illustration of this process, the behavior of the eigenvalues of the monodromy matrix associated with $\mathcal{O}_{1}$ is shown on Fig.~\ref{bifu} (for $\omega=1.67$, similar to the case $\omega=1.42$). The two eigenvalues, initially conjugated on the unity circle (i.e. $\mathcal{O}_{1}$ is elliptic), encounter a period doubling bifurcation when they reach
$1$, and leave the unity circle~: $\mathcal{O}_{1}$ has turned
hyperbolic. Then they are of the form $(\lambda,1/\lambda),\lambda\in\mathbb{R}^*$.
Eventually, the phenomenon will revert, the eigenvalues going back
on the circle and $\mathcal{O}_{1}$ to ellipticity. For $\epsilon$ between $0.4$ and $0.9$,   the orbits $\mathcal{O}_{1}$, $\mathcal{O}_{2}^{+}$ and $\mathcal{O}_{2}^{-}$ are all hyperbolic, acknowledging a complete mixing. Then, increasing further $\epsilon$, the orbit $\mathcal{O}_{2}^{+}$ turns elliptic after $\epsilon \approx 0.9$, and hence mixing decreases. It remains so until it merges with the hyperbolic $\mathcal{O}_{1}$, at $\epsilon \approx 1.1$, to give an elliptic $\mathcal{O}_{1}$ (parabolic at the transition), according to the scheme~:

\begin{equation} \mathcal{O}_{2}^{+}[e] + \mathcal{O}_{1}[h] \rightarrow \mathcal{O}_{1}[e]. \label{bifehe}\end{equation}

Beyond $\epsilon \approx 1.4$, the only remaining orbit, $\mathcal{O}_{1}$, stays elliptic.

For $\omega=1.67$ (see Fig.~\ref{om167}), the dynamics is more regular~: While bifurcation~(\ref{bifheh}) still occurs, resulting in an hyperbolic $\mathcal{O}_{1}$, the residue of $\mathcal{O}_{2}^{+}$ never crosses $1$, which means that the orbit stays elliptic. Around $\epsilon \approx 0.9$, it merges with the hyperbolic $\mathcal{O}_{1}$ to give an elliptic $\mathcal{O}_{1}$, according to the bifurcation~(\ref{bifehe}). Beyond this point, $\mathcal{O}_{1}$ will stay elliptic~: Full mixing cannot be achieved for such values of $\omega$.

Three other curves illustrate the behaviors at low frequency, i.e. $\omega=0.8$ and $\omega=0.58$, and at a high frequency $\omega=2.08$.

For $\omega=0.8$ (see Fig.~\ref{om08}), the bifurcation scheme is more complicated due to the presence of the orbit $\mathcal{O}_{1}^{\alpha}$ and $\mathcal{O}_{1}^{\beta}$ (resp. right and left dashed line). The orbits can either exist in an elliptic or hyperbolic way, or not exist at all. The two latter cases, combined with $\mathcal{O}_{1}$ and $\mathcal{O}_{2}^{+}$ hyperbolicity, are suitable for a fully mixing regime.

For low $\epsilon$, though $\mathcal{O}_{2}^{+}$ turns hyperbolic very soon (at $\epsilon \approx 0.05$), $\mathcal{O}_{1}^{\beta}$ is elliptic until $\epsilon \approx 0.2$, when it disappears. However, because of the ellipticity of $\mathcal{O}_{1}$, full mixing cannot occur until it also turns hyperbolic, at $\epsilon \approx 0.3$. Then complete mixing is achieved until $\epsilon\approx 1.5$, when $\mathcal{O}_{2}^{+}$ turns back elliptic, soon followed by $\mathcal{O}_{1}$ when they merge according to the bifurcation scheme (\ref{bifehe}). Note that around $\epsilon\approx 1.6$, orbit $\mathcal{O}_{1}^{\alpha}$ appears, but full mixing is no more possible because of $\mathcal{O}_{1}$ ellipticity.

Then, for $\omega=0.58$, the bifurcation scheme is more intricate due to the fact that $\mathcal{O}_{1}^{\beta}$ turns hyperbolic. For low $\epsilon$, despite $\mathcal{O}_{2}^{+}$ and $\mathcal{O}_{1}$ soon turn hyperbolic (at $\epsilon\approx 0.2$ for the latter), $\mathcal{O}_{1}^{\beta}$ remains elliptic until $\epsilon\approx 0.5$, when its residues goes above $1$~: full mixing is then achieved (see Fig.~\ref{epsom8}), until the orbit returns to ellipticity at $\epsilon\approx 0.65$. However, it disappears at $\epsilon\approx 0.7$, and then only $\mathcal{O}_{1}$ and $\mathcal{O}_{2}^{+}$ are present, in their hyperbolic form; thus mixing is achieved anew. It remains so until $\epsilon\approx 1.1$, when $\mathcal{O}_{1}^{\alpha}$ comes into play, being elliptic. Furthermore, it will also be hyperbolic (around $\epsilon\approx2$), but only when $\mathcal{O}_{1}$ has turned back to ellipticity~: mixing will not occur any longer.

For large $\omega$ ($\omega =2.08$, see Fig.~\ref{om208}), the dynamics is more regular. The orbits $\mathcal{O}_{2}^{+}$ and $\mathcal{O}_{2}^{-}$ do not exist, and the only remaining orbit, $\mathcal{O}_{1}$, never encounters any bifurcation, and stays elliptic. Complete mixing cannot be achieved in this case either.

\begin{figure}
\centering \subfigure[$\epsilon=0.8$]{\label{eps08}\includegraphics[width=8cm]{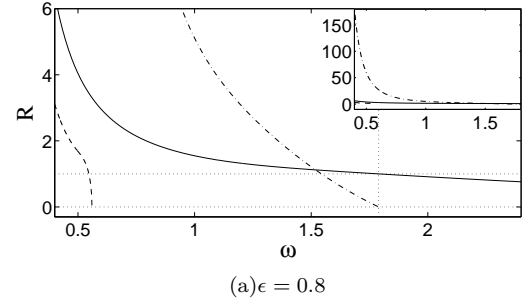}}
\\

\subfigure[$\epsilon=0.4$]{\label{eps04}\includegraphics[width=8cm]{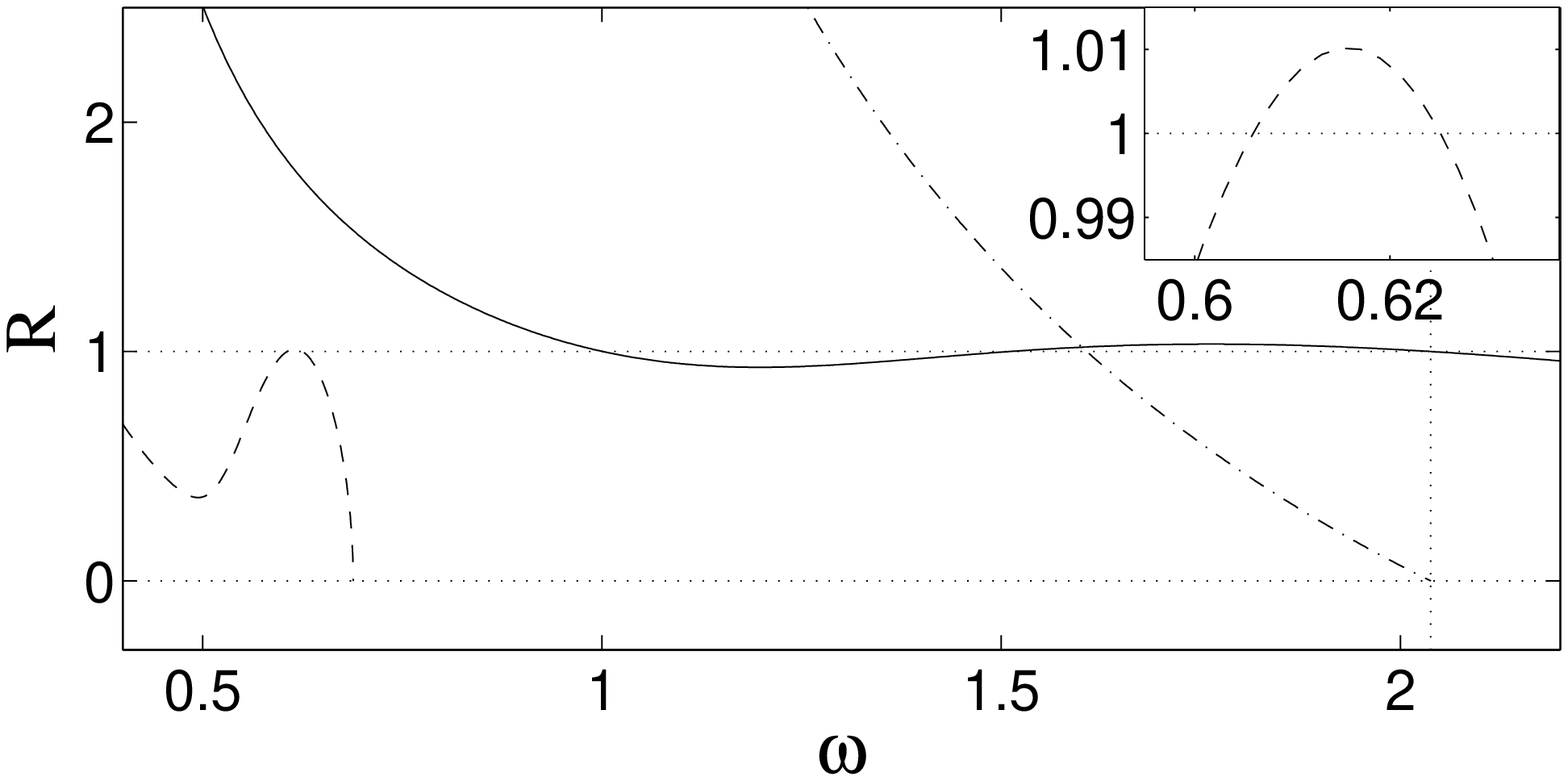}}
\\

\caption{Residue curves for the stream function~(\ref{Strmpar}) as functions of the frequency $\omega$~: The plain line corresponds to the residues of $\mathcal{O}_{1}$,
the dashed one to the residues of $\mathcal{O}_{1}^{\beta}$ and the dash-dotted
to the ones of $\mathcal{O}_{2}^{+}$ (upper curve) and $\mathcal{O}_{2}^{-}$
(lower curve). The dotted lines indicate the locations of the bifurcations.\label{epscst}}
\end{figure}

Now, instead of varying the amplitude $\epsilon$ of the forcing, we vary its frequency $\omega$ and keep $\epsilon$ constant. For $\epsilon =0.8$ (see Fig.~\ref{eps08}) and for $\omega$ below $0.55$, the orbit $\mathcal{O}_{1}^{\beta}$ is hyperbolic as well as $\mathcal{O}_{1}$ and $\mathcal{O}_{2}^{+}$~: The cell is fully mixing. Soon after $\mathcal{O}_{1}^{\beta}$ turns elliptic at $\omega \approx 0.55$, it disappears around $\omega \approx 0.57$; the remaining orbits are hyperbolic until $\mathcal{O}_{2}^{+}$ turns elliptic ($\omega \approx 1.55$). Finally, at $\omega \approx 1.8$, the elliptic $\mathcal{O}_{2}^{+}$ merges with the hyperbolic $\mathcal{O}_{1}$ (as described by the bifurcation~(\ref{bifehe})) to give an elliptic $\mathcal{O}_{1}$~: Only partial mixing is achieved.

For $\epsilon =0.4$ (see Fig.~\ref{eps04}), the situation is slightly different~: For low $\omega$, the orbit $\mathcal{O}_{1}^{\beta}$ (plain line) is present and mostly elliptic. Its residue is higher than one only in a small range of $\omega$ (around $\omega =0.61$, see the inset of Fig.~\ref{eps04}), and only this small domain of $\omega$ is suitable for complete mixing, since $\mathcal{O}_{1}$ and $\mathcal{O}_{2}^{+}$ are hyperbolic for small $\omega$. Then, for $\omega \approx 0.7$, the elliptic $\mathcal{O}_{1}^{\beta}$ disappears, and as long as $\mathcal{O}_{1}$ (dashed line) and $\mathcal{O}_{2}^{+}$ (dash-dotted line) are hyperbolic, the cell is still fully mixing. Around $\omega=1$, $\mathcal{O}_{1}$ turns elliptic and so mixing is only partial. Moreover, when it turns back to hyperbolicity for $\omega \approx 1.55$, the orbit $\mathcal{O}_{2}^{+}$ soon becomes elliptic~: The parameter range available for complete mixing is small. Finally, $\mathcal{O}_{2}^{+}$ merges with the hyperbolic $\mathcal{O}_{1}$ through the bifurcation~(\ref{bifehe}), leaving an elliptic $\mathcal{O}_{1}$~: Full mixing cannot be achieved any more.

\begin{figure}
\includegraphics[width=9cm]{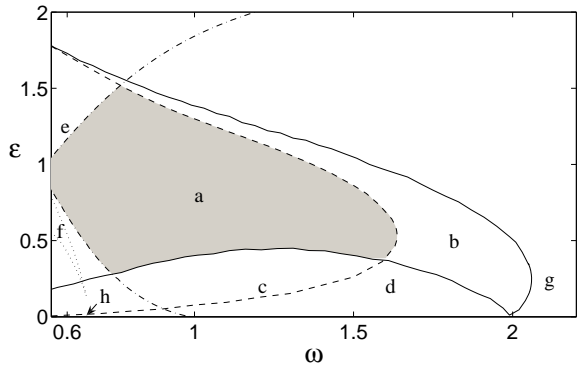}
\caption{\label{epsom} Mixing domains for the stream function~(\ref{Strmpar}) in parameter space~: The letters (a-h) refer to mixing regimes as depicted in the typical Poincar\'e sections in Fig.~\ref{eightsec}. The gray domain represents the expected mixing regime where the studied periodic orbits are all hyperbolic (or non-existent). The plain curve is associated with the orbit ${\mathcal O}_{1}$ where its residue is equal to one; the dashed curve is associated with the orbit ${\mathcal O}_{2}^{+}$; the upper dash-dotted line bounds the domain of existence of ${\mathcal O}_{1}^{\alpha}$ (it is elliptic above and does not exist below), while the lower dash-dotted line delimits the one of  ${\mathcal O}_{1}^{\beta}$ (existing below but not above). While the latter orbit is mainly elliptic, the dotted line encloses the small domain where it is hyperbolic.}
\end{figure}

Figure~\ref{epsom}
summarizes the residue study with the domain of ellipticity/hyperbolicity
of these orbits. The domains of parameters are noted with letters,
in agreement with the labeling of Fig.~\ref{eightsec}. The gray colored domain is the most suitable for complete mixing, since $\mathcal{O}_{1}$
and $\mathcal{O}_{2}^{+}$ are hyperbolic, while $\mathcal{O}_{1}^{\beta}$ and $\mathcal{O}_{1}^{\alpha}$
do not exist. The domain (f) would also be suitable; however, in this range of parameters, new orbits are born when $\epsilon$
decreases.

\section{Conclusion and perspectives}

Time-periodic perturbations are able to generate chaotic mixing in
two-dimensional channels. Optimal mixing is obtained when all regular
structures are broken by the perturbation. We have shown here that
this optimal mixing can be obtained in a small domain of phase space
by combining two strategies~: First, by constructing two transport
barriers, confining the fluid in a bounded region of the channel.
Generically, it is expected that the motion inside this bounded cell
is a mixture of regular (non-mixing) regions and chaotic (mixing)
ones. Then, using the linear stability of a few selected periodic orbits (represented
by their residues) and the identification of bifurcations, we
gave conditions on the parameters of the system (represented by the amplitude
and the frequency of the periodic forcing) such that a high mixing
occurs in the cell. We have shown that complete mixing is expected
in a large region of parameter space. The mixing properties have been a posteriori analyzed
and confirmed by computing a finite time Lyapunov map of initial conditions space. Moreover a strategy for placing an initial
drop of dye has been proposed by combining the information of the finite time Lyapunov map with a map in initial conditions space of average return times.
Finally, we have shown that the  strategy we developped is robust to several effects like truncations
of the Fourier series giving the exact shape of the perturbation,
three-dimensional effects and molecular diffusion. When  no-slip boundary conditions apply,  it has to be taken into account in the computation of the perturbation since the restored transport barrier are not robust without this, although the good mixing properties do not seem to be affected.

\bibliographystyle{apsrev}

\end{document}